\documentclass[twocolumn,showpacs,superscriptaddress,aps,prc,10pt,nofootinbib]{revtex4-1}
\usepackage[pdftex]{graphicx}
\usepackage{rotating}
\usepackage{dcolumn}
\usepackage[usenames,svgnames,table]{xcolor}
\usepackage[english]{babel}
\usepackage{ucs}
\usepackage[utf8x]{inputenc}
\usepackage{siunitx}
\usepackage[version=3]{mhchem}
\usepackage{multirow}
\usepackage{color}
\usepackage{wasysym}
\begin{document}
\sisetup{exponent-product = \cdot,per-mode = symbol,range-phrase = 
--,list-units = single,range-units = single,separate-uncertainty = 
true,table-number-alignment = center}

\preprint{}

\title{Comparison of AMD calculations with experimental data for
    peripheral collisions\\ of $^{93}$Nb+$^{93}$Nb,$^{116}$Sn at 38 MeV/nucleon}
% Force line breaks with \\

\author{S.~Piantelli}
\email{Corresponding author. e-mail: piantelli@fi.infn.it}
\affiliation{INFN Sezione di Firenze, I-50019 Sesto Fiorentino, Italy}

\author{A.~Olmi}
\affiliation{INFN Sezione di Firenze, I-50019 Sesto Fiorentino, Italy}

\author{P.~R.~Maurenzig}
\affiliation{INFN Sezione di Firenze, I-50019 Sesto Fiorentino, Italy}
\affiliation{Dipartimento di Fisica, Universit\'a di Firenze, I-50019 Sesto Fiorentino, Italy}

\author{A.~Ono}
\affiliation{Department of Physics, Tohoku University, Sendai 980-8578, Japan}

\author{M.~Bini}
\affiliation{INFN Sezione di Firenze, I-50019 Sesto Fiorentino, Italy}
\affiliation{Dipartimento di Fisica, Universit\'a di Firenze, I-50019 Sesto Fiorentino, Italy}

\author{G.~Casini}
\affiliation{INFN Sezione di Firenze, I-50019 Sesto Fiorentino, Italy}

\author{G.~Pasquali}
\affiliation{INFN Sezione di Firenze, I-50019 Sesto Fiorentino, Italy}
\affiliation{Dipartimento di Fisica, Universit\'a di Firenze, I-50019 Sesto Fiorentino, Italy}

\author{A.~Mangiarotti}
\affiliation{Instituto de F\'{i}sica da Universidade de S\~{a}o Paulo,
  05508-090 S\~{a}o Paulo, Brazil}

\author{G.~Poggi}
\affiliation{INFN Sezione di Firenze, I-50019 Sesto Fiorentino, Italy}
\affiliation{Dipartimento di Fisica, Universit\'a di Firenze, I-50019 Sesto Fiorentino, Italy}

\author{A.~A.~Stefanini}
\affiliation{INFN Sezione di Firenze, I-50019 Sesto Fiorentino, Italy}
\affiliation{Dipartimento di Fisica, Universit\'a di Firenze, I-50019 Sesto Fiorentino, Italy}

\author{S.~Barlini}
\affiliation{INFN Sezione di Firenze, I-50019 Sesto Fiorentino, Italy}
\affiliation{Dipartimento di Fisica, Universit\'a di Firenze, I-50019 Sesto Fiorentino, Italy}

%\author{A.~Buccola}
%\affiliation{INFN Sezione di Firenze, I-50019 Sesto Fiorentino, Italy}
%\affiliation{Dipartimento di Fisica, Universit\'a di Firenze, I-50019 Sesto Fiorentino, Italy}

\author{A.~Camaiani}
\affiliation{INFN Sezione di Firenze, I-50019 Sesto Fiorentino, Italy}
\affiliation{Dipartimento di Fisica, Universit\'a di Firenze, I-50019 Sesto Fiorentino, Italy}

\author{C.~Ciampi}
\affiliation{INFN Sezione di Firenze, I-50019 Sesto Fiorentino, Italy}
\affiliation{Dipartimento di Fisica, Universit\'a di Firenze, I-50019 Sesto Fiorentino, Italy}

\author{C.~Frosin}
\affiliation{INFN Sezione di Firenze, I-50019 Sesto Fiorentino, Italy}
\affiliation{Dipartimento di Fisica, Universit\'a di Firenze, I-50019 Sesto Fiorentino, Italy}

\author{P.~Ottanelli}
\affiliation{INFN Sezione di Firenze, I-50019 Sesto Fiorentino, Italy}
\affiliation{Dipartimento di Fisica, Universit\'a di Firenze, I-50019 Sesto Fiorentino, Italy}

\author{S.~Valdr\'e}
\affiliation{INFN Sezione di Firenze, I-50019 Sesto Fiorentino, Italy}

\date{\today}% It is always \today, today,

\begin{abstract}
  Experimental data concerning binary events in peripheral collisions for
  the systems $^{93}$Nb+$^{93}$Nb and $^{93}$Nb+$^{116}$Sn
  at 38 MeV/nucleon, collected with the \textsc{Fiasco} setup, are compared
  with calculations performed with the dynamic code AMD, coupled with the
  statistical code \textsc{Gemini} used as an afterburner.
  The comparison focuses on the properties of the quasiprojectile (QP)
  and on the total multiplicities of the emitted light charged particles.
  A good reproduction of the average
  mass ratio, charge $Z$ and c.m. angle of the QP is
  obtained in the examined impact parameter range (b~$\approx$ 7-12 fm).
  Concerning the light charged particles, a general agreement
  is found for the total emitted charge, while some discrepancy
  remains for the multiplicities of the various species, especially for
  the protons which are always overestimated by the calculations.
\end{abstract}

%\keywords{Suggested keywords} %Use showkeys class option if keyword
\maketitle

%--------------------------------------------------------------------

\section{Introduction}
\label{introd}

Heavy ion collisions in the Fermi energy domain (20-50 MeV/nucleon)
represent a transition regime, where the mean field plays an important
role (like at low energies), but the nucleon-nucleon collisions become
more and more relevant in determining the dynamics.
As such, heavy ion reactions at Fermi energies represent a challenge
for theoretical models, because they display a variety of phenomena
strongly dependent on the impact parameter.
In fact, in peripheral and semiperipheral reactions the
cross section is dominated by binary exit channels, with the production of
two main fragments, the quasi-projectile (QP) and the quasi-target (QT);
they may be accompanied by a significant contribution of midvelocity
emissions
\cite{Baran04,Bowman93,Montoya94,Toke95,Dempsey96,Toke96,Lukasik97,Plagnol99,Piantelli2002,Piantelli2006}.
At the other extreme, multifragmentation phenomena represent a very
important reaction channel for central collisions
\cite{Lecolley1996,Marie1997,Dagostino1996,Frankland2001A,Frankland2001B}.

Transport models are an important tool to describe these reactions
and they are able to account for many aspects, although a unified
description suitable for the whole impact parameter range, from central
to peripheral collisions, is difficult.
Such models can be divided into two main classes.
In one class there are the models based on the BUU
(Boltzmann-Uehling-Uhlenbeck) approach, which follow the evolution in
time of the phase space density: e.g., among many others, the
Stochastic Mean Field (SMF) \cite{Colonna1998} and the
Boltzmann Langevin One Body (BLOB) \cite{Napolitani2013BLOB}.
In the other class there are the models that follow the evolution of
the nucleon coordinates and momenta, namely the various flavors of
quantum molecular dynamics (QMD) models.
Very recently Zhang et al. \cite{Zhang18} performed a very
extensive theoretical comparison of different models, belonging to both
classes of BUU and QMD type, focusing on the most critical ingredient of
the transport equations, namely the collision term.
In another recent paper by Xu et al. \cite{Xumodelli}, the comparison
of several different models of the two classes was performed by
simulating the same collision, namely $^{197}$Au+$^{197}$Au, at 100 and
400 MeV/nucleon, with a fixed impact parameter of $b$=7 fm.
In particular, the rapidity distribution and the collective flow were
compared, finding a considerable spread in the outcome of the different
models.
As a consequence it is particularly useful to compare the predictions
of the models with experimental data, in order to establish more
stringent constraints.

Among the various QMD models we consider here the AMD
(Antisymmetrized Molecular Dynamics \cite{Ono92}), which is able to
give a good description of the main characteristics of heavy ion
reactions at Fermi energies.
In the literature one can find some comparisons of AMD calculations 
and experimental data, but they are mainly focused on central collisions
(see, e.g., \cite{Ono02,OnoJPC2013})
or based on inclusive data \cite{Tian2018}.
For example in Ref. \cite{Tian2018} it was shown that some properties
(like angular distributions, energy spectra and production cross section)
of intermediate mass fragments (IMF) observed in inclusive measurements
for the system $^{12}$C+$^{12}$C at 95 MeV/nucleon are well reproduced
by AMD followed by \textsc{Gemini++} \cite{Charity10} as an afterburner.

In this paper the focus is on the QP properties and on the light
charged particles (LCP) produced in peripheral and semiperipheral
heavy ion collisions for the systems $^{93}$Nb+$^{93}$Nb and
$^{93}$Nb+$^{116}$Sn at 38 MeV/nucleon.
It will be shown that for these systems the AMD model, coupled with
\textsc{Gemini} \cite{Charity88,Charity90,Charity10}, is well suited to
describe the average characteristics of the projectile remnants in binary
collisions, for the upper half of the impact parameter range
(from about  0.5~b$_{\mathrm{graz}}$ to b$_{\mathrm{graz}}$).

The experimental data used here have already been the subject of other
papers \cite{Mangiarotti2004,Piantelli2006,Piantelli2007,Piantelli2008}
that were mainly focused on the properties of LCPs and IMFs.
On the basis of the results published therein, it will be shown that
not only the QP properties, but also the experimental total particle
multiplicities in peripheral and semiperipheral binary collisions are
reasonably well reproduced by the calculations, with the possible exception
of protons, which appear to be somewhat overestimated.

%--------------------------------------------------------------------

\section{Experimental setup}
\label{setup}

The experimental data were collected with the \textsc{Fiasco} setup,
which is described in detail elsewhere \cite{Bini03}.
Here only the main characteristics are briefly recalled.
The \textsc{Fiasco} setup consisted of different types of detectors.
There was a shell of 24 position sensitive Parallel Plate Avalanche
Detectors (PPADs) to measure the velocity vectors of heavy ($Z > 9$)
fragments with high efficiency \cite{Bini03} and low energy thresholds
($\sim$ 0.1 MeV/nucleon), so that they were able to detect also the
low-energy QT.
The angular coverage was about 70\% of the forward hemisphere, 
from 0.2$^{\circ}$ up to about 90$^{\circ}$.
In the polar range 0.5$^{\circ}$ -- 6$^{\circ}$, behind the 6 most forward
PPADs, 96 $\Delta E - E$ Silicon telescopes (with a thickness of
200 $\mu$m for the first layer and of 500 $\mu$m for the second one)
were devoted to the measurement of charge and energy of the QP.
Therefore, when the QP was detected in coincidence by a PPAD and a
Silicon telescope behind it, it was possible to obtain also the mass
of the QP by means of the information on its energy and time-of-flight.
The setup was completed by 182 three-layer phoswich telescopes, covering
about 30\% of the forward hemisphere, aimed at identifying the mass of
LCPs (p, d, t and $\alpha$) and the charge of heavier fragments
(in the range $Z=1\sim26$) and measuring their time-of-flight.

%--------------------------------------------------------------------

\section{AMD model}
\label{AMD-model}

The AMD model is described in detail elsewhere
\cite{Ono92,Ono99,OnoJPC2013,Ikeno16},
so only the main features are briefly recalled here.
AMD is a transport model which describes the time evolution of a
system of nucleons, by depicting the state of the system at each time step
as a Slater determinant of Gaussian wave packets.
The time evolution is achieved by applying the time-dependent
variational principle, thus obtaining an equation of motion governed
by an Hamiltonian that describes the mean field contribution by means
of an effective interaction.
Such an interaction is based, in the present case, on the
Skyrme-force parametrization SLy4 of \cite{ChabanatSLy4} with
a soft symmetry energy (slope parameter $L$ = 46 MeV),
while the normal density term, $S_0$, has the
standard value of 32 MeV \cite{OnoJPC2013}.
A stiff symmetry energy ($L$ = 108 MeV) can be obtained by changing
the density dependent term in the SLy4 force \cite{Ikeno16}.
Unless otherwise stated, the stiff symmetry energy is used in this paper.

In the present work, we employ a new method for two-nucleon collisions
based on the test particles which are randomly generated at every time step.
The test particles are the samples taken from the exact one-body Wigner
function defined for the AMD wave function of antisymmetrized Gaussian
wave packets.
See Appendix C of Ref.~\cite{Ikeno16} for the method to generate
test particles.
The attempt of a collision between two test particles is judged by a
geometrical condition as in many other
transport codes (see e.g.~Ref.~\cite{Zhang18}).
A possible benefit of this new method is that the collisions will reflect
the exact density distribution in contrast to the previous method that
employs the so-called physical coordinates \cite{Ono92ptp} which can
represent the phase-space distribution only approximately.
In the new method, when it is decided that two test particles
$(\boldsymbol{r}_1,\boldsymbol{p}_1)$ and
$(\boldsymbol{r}_2,\boldsymbol{p}_2)$ collide, a collision
is still performed by changing the momenta of the two physical coordinates
$(\boldsymbol{R}_{k_1},\boldsymbol{P}_{k_1})$ and
$(\boldsymbol{R}_{k_2},\boldsymbol{P}_{k_2})$ that are
associated with the two colliding test particles.
The final momenta $\boldsymbol{P}'_k$ ($k=k_1, k_2$) are allowed by
the Pauli principle when
$\nu |\boldsymbol{R}_k - \boldsymbol{R}_j|^2 +
|\boldsymbol{P}'_k - \boldsymbol{P}_j|^2 / (4 \hbar^2 \nu) < 1.46^2$ is
satisfied for all $j$ ($\ne k$) with the same spin-isospin state as $k$.
Here $\nu$ is the width parameter with the usual value of
0.16 $\text{fm}^{-2}$ \cite{Ono96,Ikeno16}.
The existence of the backward transformation from the physical coordinates
to the variables of an AMD wave function is also required \cite{Ono92ptp}.
The transition probability depends on the {\em in-medium}
nucleon-nucleon cross section, which can be considered, within some
limits, as a free parameter of the model.
In the version of the code used in this work, the parametrization
introduced in \cite{Coupland11} was adopted, namely
\begin{equation}
  \sigma=\sigma_0 \tanh(\sigma_\mathrm{free}/\sigma_0),
  ~~~\mathrm{with} ~~\sigma_0=y~\rho^{-2/3}
  \label{sigmaNN}
\end{equation}
where $\rho$ is the nuclear density and $y$ a screening parameter.
With the value  $y$ = 0.85 proposed in \cite{Coupland11}, the other
model parameters, including that for the Pauli blocking,
have been chosen so as to approximately reproduce the fragment charge
distribution in the central Xe + Sn collisions at 50 and 32 MeV/nucleon
and to have the degree of stopping $R=0.55$ at 50 MeV/nucleon and $R=0.62$
at 32 MeV/nucleon which may be compared to the experimental data in
Ref.~\cite{Lopez2014}.
The calculated stopping variable is defined for the transverse and
longitudinal kinetic-energy components, $E_i^\perp$ and $E_i^\parallel$,
in the c.m.\ frame by
$R=(\sum_i E_i^\perp)/(2\sum_i E_i^\parallel)$,
where the summation is for all the light charged particles and heavier
fragments produced in all the calculated central events.
In the present work, the study of the dependence on the in-medium
cross section is extended to peripheral collisions at Fermi energies
by testing also lower values of $y$
% (0.60 and 0.42),
corresponding to a larger reduction of the in-medium cross section.

When a two-nucleon collision has occurred and the physical coordinates
have been updated to $\boldsymbol{P}'_{k_1}$ and $\boldsymbol{P}'_{k_2}$,
we may virtually consider a similar scattering of the two test particles to
the final momenta $\boldsymbol{p}'_1$ and $\boldsymbol{p}'_2$ which contain
physical fluctuations.
In the present work, we turned on an option to incorporate these
fluctuations into the dynamics.
When a wave packet $k$ is emitted at a later time, the fluctuation
$\Delta\boldsymbol{p}'_k$ of its most recent collision is added to
the momentum of the nucleon $k$.
This is a simplified way of introducing wave packet splitting, and it
should influence the energy spectra of emitted particles.
The energy and momentum conservation laws are taken into account in a
similar way to Ref.~\cite{Ono96}.
When a cluster is emitted, the sum
of the fluctuations $\Delta\boldsymbol{p}'_k$ of its nucleons
is added to the c.m.\ motion of the cluster.

The cluster correlations are explicitly taken into account by allowing
each of the scattered nucleons to form a light cluster such as a deuteron,
triton and $\alpha$ particle.
The method is the same as that employed in Refs.~\cite{Ikeno16,Tian2018},
except that a new method is adopted in the present work to suppress the
cluster correlation in nuclear medium.
The probability of attaching a nucleon $i$ to one of the scattered nucleons
$k$ (or a subcluster $k$) is reduced by multiplying a factor $1-0.3f$,
where $f$ is an approximate Wigner function, with the contribution from
$i$ excluded, at the phase-space point of the center of mass of $i$ and $k$.
The method of binding several clusters to form light nuclei (Li, Be, etc.)
is almost the same as that of Ref.~\cite{Tian2018}.
However, we here choose a stricter condition for binding than in
Ref.~\cite{Tian2018} so that the chance of binding several clusters is reduced.
The necessary conditions for a pair of
clusters to be linked now include
that their relative kinetic energy should satisfy
$\frac12\mu V_{\text{rel}}^2<10$ MeV and that each of them is one of the three
closest clusters of the other.
For the energy conservation, we adjust the relative momentum of the bound
light nucleus and the third cluster that has the minimum value of a measure
$(r+7.5\,\text{fm})(1.2-\cos\theta)/\mathop{\mathrm{min}}(\epsilon_\parallel,\,5\,\text{MeV})$
as in Ref.~\cite{Tian2018}.
However, this measure is divided by a factor 2 in the present work for a
candidate cluster that is in a light nucleus already bound at a former time,
and thus light nuclei are favored as the third cluster for the energy
conservation.

The time evolution of the AMD calculation was usually stopped at a time
(from now on called ``switching time'' and indicated as $t_{\mathrm{sw}}$)
of 500 fm/c, which was verified to be a reasonable time to assume that
the dynamics of the collision has already established the final
partitions.
Many tests were done also stopping the AMD calculation at values of
$t_{\mathrm{sw}}$ from 200 up to 10000 fm/c
to verify a possible
dependence of the obtained results on the choice of the switching time.
In this paper, each AMD calculation
usually consisted of 7000 or more events, except for the case with
$t_{\mathrm{sw}}$ = 10000 fm/c where only 2500 events were generated,
due to the extremely long computation time.
The impact parameters were distributed in a triangular shape between
0 and 13 fm, a value that is slightly larger than the grazing
impact parameter of the collisions (about 12.3 and 12.6 fm
for $^{93}$Nb+$^{93}$Nb and $^{93}$Nb+$^{116}$Sn, respectively).

\begin{figure*}[ht]       % FIG.1
\centering
\includegraphics[width=\textwidth]{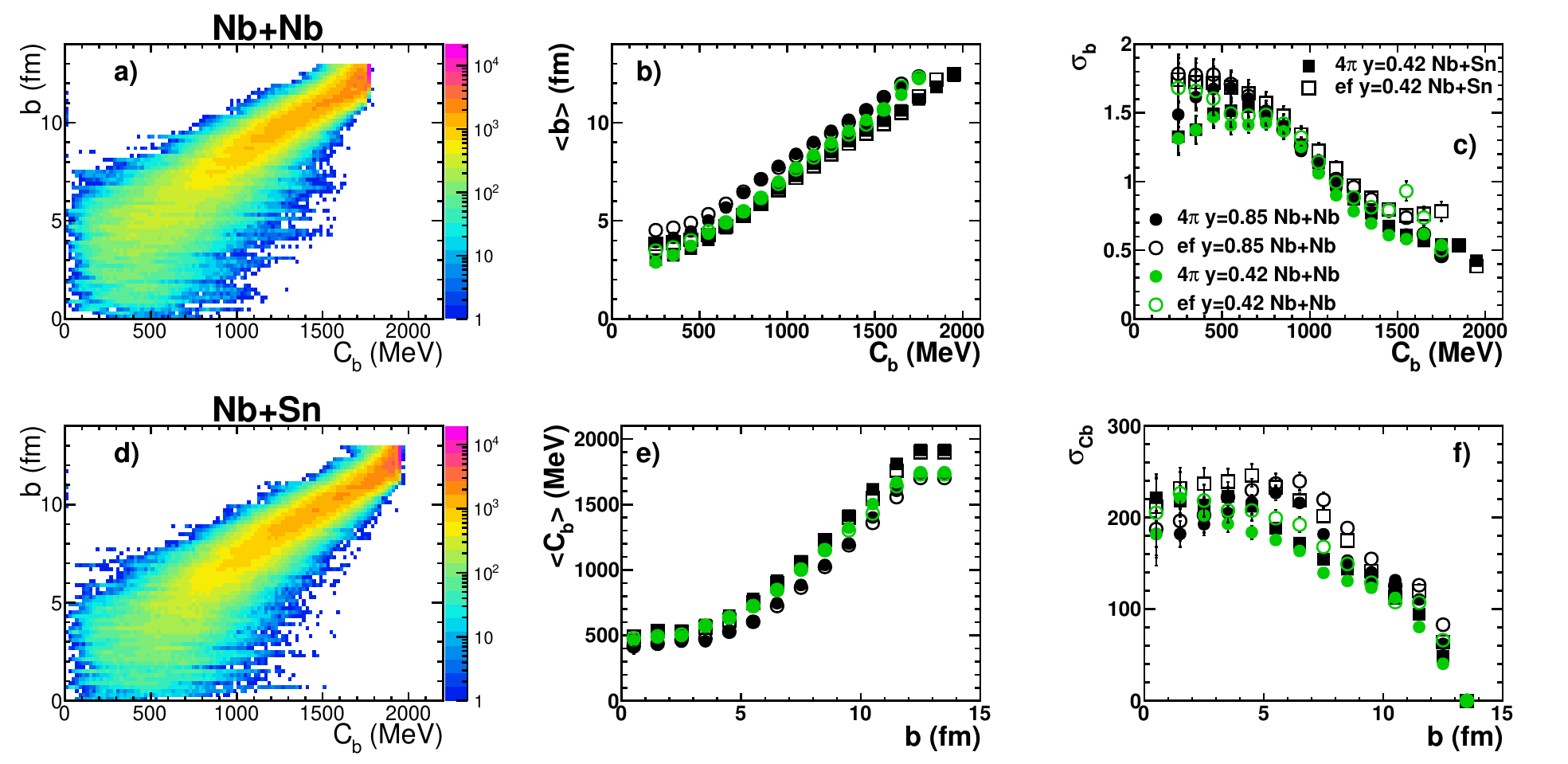}
\caption{(Color online) Results of AMD plus \textsc{Gem++} calculations,
  with $t_{\mathrm{sw}}$ = 500 fm/c.
  Two-dimensional correlations $b$ vs. $C_b$ for $4\pi$ data of
  (a) $^{93}$Nb+$^{93}$Nb (screening parameter $y$ = 0.85)
     and (d) $^{93}$Nb+$^{116}$Sn ($y$ = 0.42);
  (b) mean values and (c) standard deviations
    of the $b$ distribution as a function of $C_b$;
  (e) mean values and (f) standard deviations
    of the $C_b$ distribution as a function of $b$.
    Full and open symbols are for $4\pi$ and filtered results, respectively.
    Circles refer to $^{93}$Nb+$^{93}$Nb (black for $y$ = 0.85 and green for
    $y$ = 0.42) and squares to $^{93}$Nb+$^{116}$Sn.
}
\label{bparb}
\end{figure*}

The primary fragments produced by AMD are rather excited,
so that it is necessary to let them decay before comparing the calculated
results with the experimental data.
Therefore, 100 secondary events were generated for each AMD event
by means of a statistical afterburner.
The masses and charges of the fragments produced by a single
AMD event, as well as their excitation energies and angular momenta,
are the input parameters for the following afterburner.
For each replica of the same AMD event, the reaction plane was
axially rotated at random, in a coherent way for all the reaction products.
In this paper the statistical code \textsc{Gemini} was used as an afterburner,
both in its Fortran90 \cite{Charity88} and C++ version \cite{Charity10}
with the standard parameters \cite{Gemini-par}.
In fact the results of the two versions of the statistical code present some
differences, which are found to be appreciable
for the multiplicities of the emitted LCPs.
In the following, for brevity, the two versions of \textsc{Gemini}
will be addressed as \textsc{Gem90} and \textsc{Gem++}, respectively.

The so produced secondary events were then filtered with a software
replica of the experimental setup, keeping into account both the
geometrical coverage and the detection thresholds of the PPADs.
Concerning the LCPs and IMFs, no filtering was performed because their
multiplicities were compared with the published data of Ref.
\cite{Piantelli2006}, which had already been corrected for the geometrical
coverage of the phoswich telescopes\footnote{The reliability
  of the correction can be appreciated from Fig.~1
  of \cite{Mangiarotti2004}, where the total charge in the forward c.m.
  hemisphere comes close to 41 (the charge of the projectile), 
  with a small average deficit of about half a charge unit.
}.
The experimental effects can be appreciated from the comparison of the
filtered events with those produced directly by the calculations,
which will be shortly called ``4$\pi$ events'' in the following.

One point to be noted is that, while the AMD calculation takes into proper
account the mutual Coulomb repulsion of all the reaction products,
\textsc{Gemini} makes each product decay by its own, without further
acceleration.
This means that when using short values of $t_{\mathrm{sw}}$,
the final secondary fragments will lack the full Coulomb reacceleration
and their velocity vectors will not have the proper asymptotic values.
This affects both the emission angles and kinetic energies, which
are therefore expected to be somewhat too low.
For example,
for $t_{\mathrm{sw}}$ = 200 (500) fm/c the total kinetic energy in the c.m.
frame of two Nb-like fragments will be short of about 45 (20) MeV with
respect to the asymptotic value of fully accelerated fragments,
corresponding to a kinetic energy deficit of about 2.5 \% (1.1 \%) for
quasi-elastic events.
This minor effect has been taken into account by applying a small
multiplicative correction to the $C_b$ scale (see next section)
before comparing calculations performed with different values of the
switching-time parameter.

%--------------------------------------------------------------------

\section{Results}
\label{results}

\subsection{Event sorting}
\label{e-sorting}

Peripheral and semiperipheral collisions have been experimentally
selected by requiring
that only two heavy fragments ($Z >$ 9) are detected in the PPADs.
The selection of binary events was implemented by rejecting
those events that severely violate the binary kinematics, based
on the relative angle $\alpha$ between the c.m. velocities of the two
detected fragments ($\cos(\alpha)\leq -0.8$) and on the difference
between their azimuthal angles
($|\varphi_{\,\mathrm{QP}}-\varphi_{\,\mathrm{QT}}|=180^\circ \pm 20^{\circ}$).
This selection will be referred to as
``collinearity condition'' in the following.
Of the two heavy fragments, the forward-emitted one (in the c.m. frame)
is assumed to be the QP, the other the QT.
The same selections are applied to the calculated events.

Since the impact parameter is not accessible to experiments, it is
necessary to find another observable that allows a fair comparison
between calculated and experimental events as a function of the
centrality of the collision.
In the present case we introduce the variable $C_b$ defined as
\begin{equation}
  C_b = \frac{1}{2}~ M~ v_{\mathrm{QP}}^{\,\mathrm{cm}}~ v_{\mathrm{QT}}^{\,\mathrm{cm}},
\label{Cb}
\end{equation}
where $v_{\mathrm{QP(QT)}}^{\,\mathrm{cm}}$ is the
secondary velocity of the QP (QT) in the
c.m. reference frame and $M$ is the total mass of the system.
The relationship of $C_b$ with the more common TKE
  (Total Kinetic Energy) will be explained later in this section.

The correlation between $C_b$ and the impact parameter $b$ was
studied by means of
the events produced by AMD plus \textsc{Gemini} and is shown
in Fig. \ref{bparb}.
Panels (a) and (d) display the two-dimensional correlations
$b$ vs. $C_b$ (in  $4\pi$) for the systems $^{93}$Nb+$^{93}$Nb
and $^{93}$Nb+$^{116}$Sn, respectively.
For a more quantitative analysis, the average impact parameter
$\langle b\rangle$ and the standard deviation of the $b$ distribution
are shown as a function of $C_b$ in panels (b) and (c),
respectively, for both reactions.
Panels (e) and (f) show the same information the other way around, namely
the mean $\langle C_b\rangle$ and the standard deviation of the $C_b$
distribution, respectively, this time as a function of $b$.
Full symbols refer to calculations for binary events in $4\pi$.
Full circles are for the systems $^{93}$Nb+$^{93}$Nb (black for
a value $y$ = 0.85 of the screening parameter defined in Eq.~(\ref{sigmaNN})
and green for $y$ = 0.42)
and full squares refer to the system $^{93}$Nb+$^{116}$Sn ($y$ = 0.42).
From panels (b) and (e), one observes that there is a
good average correlation between $C_b$ and $b$, from grazing collisions
down to $C_b \approx$ 500 MeV, or to $b \approx$ 5 fm
and that the sensitivity to the screening parameter $y$ is negligible.
Panels (c) and (f) show also that the width of the correlation
becomes increasingly wide with increasing centrality 
and below $C_b \approx$ 1000 MeV the estimate of $b$ has an uncertainty
of the order of $\pm$ 1 fm or greater.
In the same figure, the open symbols show the negligible effect of the
experimental filter, which requires that there are only two
{\em detected} fragments with $Z > 9$ and that they must
additionally satisfy the ``collinearity condition''.
This latter request helps rejecting ternary (or higher multiplicity)
events that {\em appear} to be
binary just because only two fragments passed the experimental filter.

\begin{figure*}[htpb]     % FIG.2
\centering
\includegraphics[width=\textwidth]{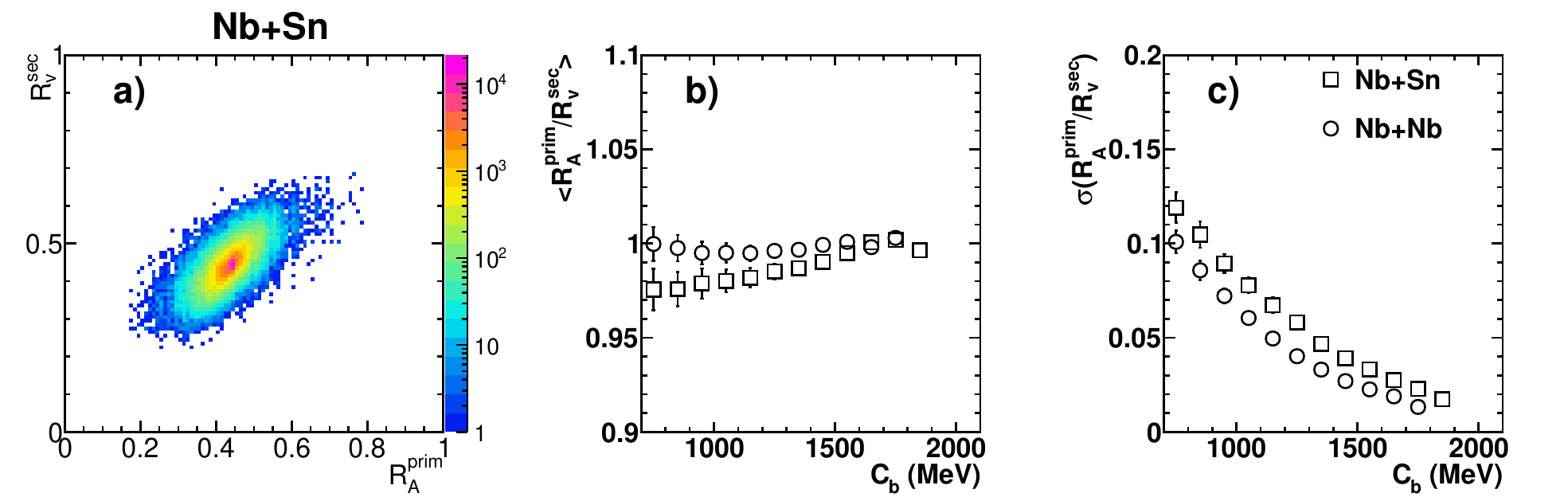}
\caption{(Color online) Results of AMD plus \textsc{Gem++} calculations,
   with $t_{\mathrm{sw}}$ = 200 fm/c:
  (a) two-dimensional distribution of $R_v$ vs. $R_A$ for semiperipheral
    binary events with $C_b > $ 700 MeV,
    in the system $^{93}$Nb+$^{116}$Sn
    ($y$ = 0.42) at 38 MeV/nucleon (events filtered
    with the efficiency of the experimental setup);
  (b) average ratio $\langle R_A/R_v \rangle$ as a function of $C_b$, for both
    reactions $^{93}$Nb+$^{93}$Nb (circles) and $^{93}$Nb+$^{116}$Sn (squares); 
  (c) standard deviation of the distribution of $R_A/R_v$
    as a function of $C_b$.
}
\label{rap}
\end{figure*}

In a previous paper \cite{Piantelli2006}, we correlated the impact
parameter $b$ with another observable, $TKEL$ (Total
Kinetic Energy Loss), obtained with a kinematic coincidence method
(KCM \cite{KCM}). 
As shown in Fig. 15 of Ref. \cite{Piantelli2006}, such a correlation was
determined with two methods that gave similar results:
i) by direct integration of the experimental cross section
  from grazing collisions downwards, and
ii) by means of a transport code based on molecular dynamics \cite{Lukasik}.
By definition, $TKEL = E_{\mathrm{cm}}-TKE$, where
$E_{\,\mathrm{cm}}=\mu \epsilon_{\mathrm{beam}}$ is the c.m. kinetic energy in the
entrance channel \footnote{$E_{\mathrm{cm}}$ =1767 (1930) MeV for
  $^{93}$Nb+$^{93}$Nb ($^{116}$Sn) at 38\,A MeV.}
($\mu$ is the reduced mass and $ \epsilon_{\mathrm{beam}}$ the energy-per-nucleon
of the beam) and $TKE$ is the c.m. Total Kinetic Energy of QP and QT.

Within classical kinematics and in the ideal case of a true binary collision
with both $C_b$ and $TKE$ built from the c.m. primary velocities
(i.e., before statistical de-excitation), these two quantities would exactly
coincide.
In the present case,  we prefer to analyze and classify both the experimental
and the calculated quasi-binary events in terms of $C_b$, because it involves
only the secondary velocities, which are experimentally available quantities.
For what concerns the LCP multiplicities, the experimental data are taken
from Ref. \cite{Piantelli2006} and for the sake of comparison with the
calculations, also those experimental data will be presented in terms
of $C_b$, assuming that -- at least for (semi)peripheral collisions --
$C_b$ can be estimated from TKEL by means of the conversion
$C_b \approx TKE = E_{\mathrm{CM}}-TKEL$.

The comparison will be restricted to the range $C_b >$ 700 MeV, one of the
reasons (possibly the main one) being that, below this value, 3-body events 
are quite abundant in the experiment and rather scarce in the calculations.
As a consequence, any comparison both of the QP properties and of the
LCP multiplicities below $C_b$ = 700 MeV might become increasingly unreliable.
%--------------------------------------------------------------------

\subsection{QP properties}
\label{QP}

A commonly used
way to present the gross features of binary collisions at low and
intermediate energies is by means of the so-called ``diffusion plot''
(i.e. the correlation between $TKE$ and the mass of the QP) and
``Wilczynski plot'' (i.e. the correlation between $TKE$ and the c.m.
polar angle of the QP).
In previous papers \cite{CasiniPRL97,CasiniEPJ2000} concerning data
collected with a similar setup but at lower beam energies, the primary
(or pre-evaporative)
mass of the QP and its primary c.m. polar angle
were estimated by means of the KCM.
Since the kinematic method is based on the assumption of a binary reaction
(with primary masses of QP and QT adding up to the total mass of the system),
this procedure of estimating primary quantities loses its validity
at Fermi energies when
other reaction channels, such as the midvelocity emissions, become
important\,\footnote{However it was shown in \cite{Piantelli2006} that $TKEL$
  can still be used as an estimator of the centrality of the collision.}. 

Due to these drawbacks, it is better to rely on directly measured quantities.
The \textsc{Fiasco} setup measured the secondary velocity of the QP,
but this does not allow one to obtain the QP secondary mass, except for the
small number of events in which the QP was detected in coincidence
by one PPAD and a Si telescope behind it \cite{Bini03}.
However, using the secondary c.m. velocities of QP and QT, one can
build the ratio
\begin{equation}
  R_v =  \frac{v_{\mathrm{QT}}^{\,\mathrm{cm}}}{v_{\mathrm{QP}}^{\,\mathrm{cm}}+v_{\mathrm{QT}}^{\,\mathrm{cm}}},
\label{R}
\end{equation}
which was used in \cite{Piantelli2008} as an estimator of the
ratio between the QP primary mass at separation
and the sum of the primary masses of QP and QT
\begin{equation}
  R_A =  \frac{A_{\mathrm{QP}}}{A_{\mathrm{QP}}+A_{\mathrm{QT}}},
\label{RA}
\end{equation}
as far as the reaction can be considered binary.

A check of this assumption is presented in Fig. \ref{rap} using the
events produced by the AMD plus \textsc{Gem++} calculation.
Panel (a) shows the correlation between $R_v$ (calculated with
secondary velocities, i.e., after the afterburner) and $R_A$ (calculated
with the masses produced by AMD before applying the afterburner)
for $C_b >$ 700 MeV in the asymmetric system $^{93}$Nb+$^{116}$Sn.
The switching time is $t_{\mathrm{sw}}$= 200 fm/c, to be sure that the
masses of QP and QT are close to the primary values they had at separation.
The events, which are filtered with a software replica of the setup,
satisfy the collinearity condition and are binary both at the primary
and secondary level.
The figure shows a clear correlation, peaked around 0.445
(the mass ratio of the colliding system) for both variables.

To be more quantitative, Fig. \ref{rap} presents, again as a
function of $C_b$, also (b) the average ratio $\langle R_A/R_v \rangle$
and (c) its standard deviation, for both reactions
$^{93}$Nb+$^{93}$Nb (circles) and $^{93}$Nb+$^{116}$Sn (squares).
In both cases, the ratio $\langle R_A/R_v \rangle$ stays very close
to 1 at all values of $C_b$.
While this fact is just a trivial consequence of the system symmetry
in $^{93}$Nb+$^{93}$Nb, it is not so for the asymmetric collision
$^{93}$Nb+$^{116}$Sn.
There one observes that the ratio starts at about 1 for very
peripheral collisions and decreases with increasing centrality.
However the deviation from 1 is small ($\apprle 2 \%$),
so that  $R_v \approx R_A$ appears as a good approximation.
Within the model, one can check the origin of this decrease.
It is found that $\langle R_v \rangle$ remains close to 0.445 for all
values of $C_b$ and independently of the chosen value of $t_{\mathrm{sw}}$.
On the contrary, if the quantity $\langle R_A \rangle$ of Eq. (\ref{RA})
is built with the masses delivered by AMD at $t_{\mathrm{sw}}$, this mass ratio
deviates from the entrance value (0.445) towards lower values and
this deviation increases with increasing $t_{\mathrm{sw}}$.
This fact suggests that while the masses of QP and QT
become lighter and lighter, the total mass of the particles they emit
is not proportional to their initial mass.
Therefore, besides being the only one experimentally available,
$\langle R_v \rangle$ appears to be a rather good
estimator of the true primary masses at separation.
The standard deviation of $R_A/R_v$ has a similar behavior in both
reactions: it is very narrow in peripheral collisions and
then monotonically widens with increasing centrality.

\begin{figure*}[htpb]     % FIG.3
\centering
\begin{tabular}{c}
\includegraphics[width=\textwidth]{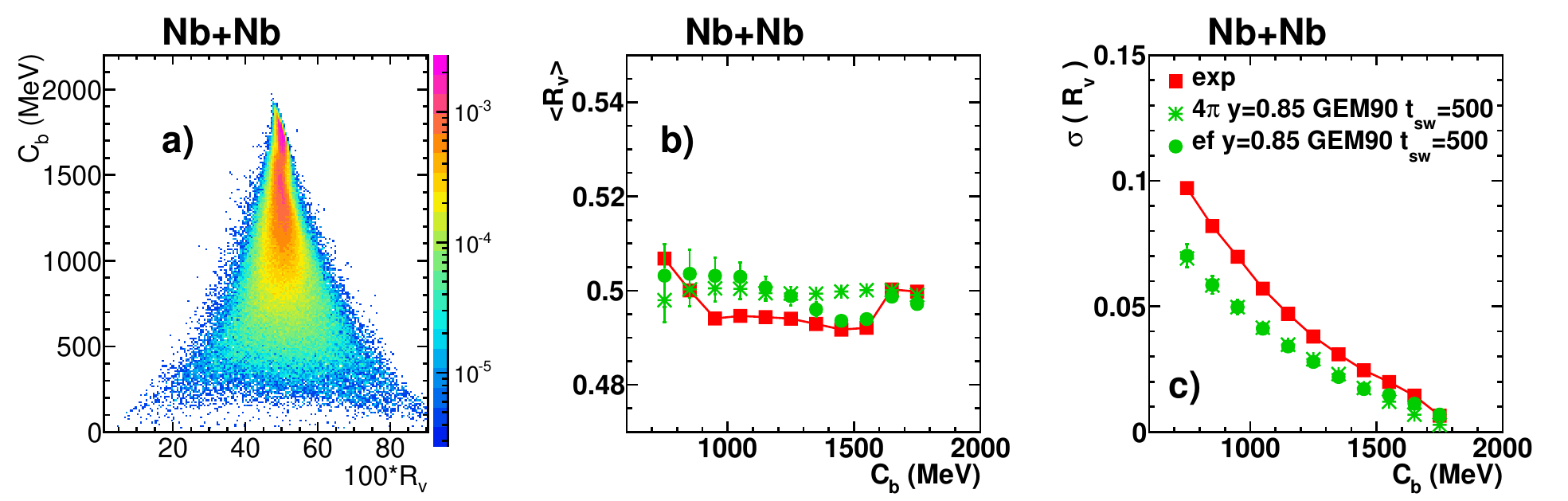}\\
\includegraphics[width=\textwidth]{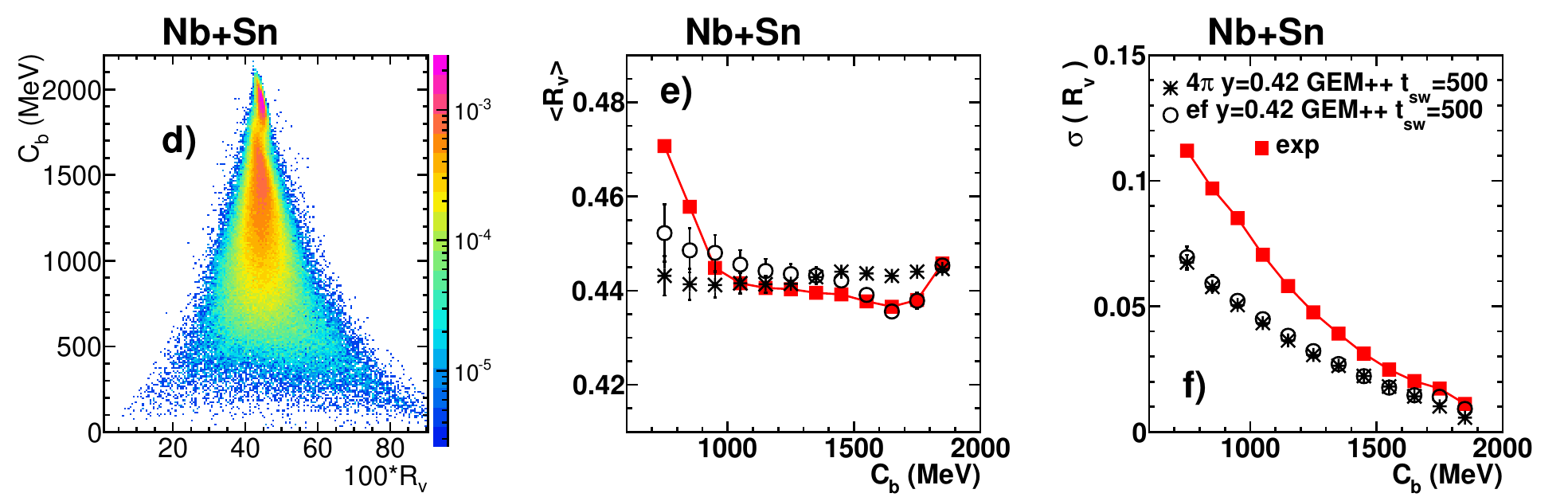}\\
\end{tabular}
\caption{(Color online) Diffusion plot. Results of AMD plus \textsc{Gemini}
  calculations for the reactions $^{93}$Nb+$^{93}$Nb (upper panel)
  and $^{93}$Nb+$^{116}$Sn (lower panels) at 38 MeV/nucleon.
  (a) Correlation $(100 R_v)$ vs. $C_b$ with model
  parameters $y$=0.85, $t_{\mathrm{sw}}$= 500 fm/c and \textsc{Gem++};
  filtered results.
  (b) Mean of the $R_v$ distribution vs. $C_b$.
  (c) Standard deviation of the $R_v$ distribution vs. $C_b$.
  (d) Same as panel (a) but for $^{93}$Nb+$^{116}$Sn with
  parameters $y$=0.42, $t_{\mathrm{sw}}$= 500 fm/c and \textsc{Gem++};
  filtered results.
  (e) Same as panel (b) but for $^{93}$Nb+$^{116}$Sn.
  (f) Same as panel (c) but for $^{93}$Nb+$^{116}$Sn.
Full (red) squares are the experimental data.
Other symbols are for calculated results; they refer to filtered events,
except for asterisks, which are in $4 \pi$.}
\label{diffusion}
\end{figure*}

\begin{table*}[htpb]    % Table I
  \begin{ruledtabular}
  \renewcommand{\arraystretch}{1.1}

\begin{tabular}{clccccccccccp{0.1mm}cccc}

\multicolumn{2}{c}{\multirow{2}{*}{}} &
\multicolumn{10}{c}{\bf Nb + Nb} & &\multicolumn{4}{c}{\bf Nb + Sn} \\ [1mm]

\multicolumn{1}{c}{} &
\multicolumn{1}{l}{$t_{\mathrm{sw}}$ (fm/c) =} &
\multicolumn{2}{c}{~~200~~} &
\multicolumn{2}{c}{~~500~~} &
\multicolumn{2}{c}{~1000~~} &
\multicolumn{2}{c}{~2500~~} &
\multicolumn{2}{c}{10000~~} & &
\multicolumn{2}{c}{~~200~~} &
\multicolumn{2}{c}{~~500~~} \\[1mm]

\cline{3-4} \cline{5-6} \cline{7-8} \cline{9-10} \cline{11-12}
                                   \cline{14-15} \cline{16-17}

\multicolumn{1}{c}{} &
\multicolumn{1}{c}{} &
\multicolumn{1}{c}{$Q_R$} &
\multicolumn{1}{c}{$\Sigma_R$} &
\multicolumn{1}{c}{$Q_R$} &
\multicolumn{1}{c}{$\Sigma_R$} &
\multicolumn{1}{c}{$Q_R$} &
\multicolumn{1}{c}{$\Sigma_R$} &
\multicolumn{1}{c}{$Q_R$} &
\multicolumn{1}{c}{$\Sigma_R$} &
\multicolumn{1}{c}{$Q_R$} &
\multicolumn{1}{c}{$\Sigma_R$} & &
\multicolumn{1}{c}{$Q_R$} &
\multicolumn{1}{c}{$\Sigma_R$} &
\multicolumn{1}{c}{$Q_R$} &
\multicolumn{1}{c}{$\Sigma_R$} \\

\multicolumn{1}{c}{$y$} &
\multicolumn{1}{c}{} &
\multicolumn{1}{c}{(\%)} &
\multicolumn{1}{c}{(\%)} &
\multicolumn{1}{c}{(\%)} &
\multicolumn{1}{c}{(\%)} &
\multicolumn{1}{c}{(\%)} &
\multicolumn{1}{c}{(\%)} &
\multicolumn{1}{c}{(\%)} &
\multicolumn{1}{c}{(\%)} &
\multicolumn{1}{c}{(\%)} &
\multicolumn{1}{c}{(\%)} & &
\multicolumn{1}{c}{(\%)} &
\multicolumn{1}{c}{(\%)} &
\multicolumn{1}{c}{(\%)} &
\multicolumn{1}{c}{(\%)} \\ [1mm]

\cline{1-17}\\ [-3mm]

\multirow{2}{*}{0.85}
& \multicolumn{1}{l}{\textsc{Gem++}}
 & 1.0 & 33 & 0.9 & 31 & 0.8 & 30 & 0.7 & 29 & 0.9 & 30 && --& --& --& --\\
& \multicolumn{1}{l}{\textsc{Gem90}}
 & 0.9 & 30 & 0.9 & 28 & 0.7 & 27 & 0.7 & 27 & 0.9 & 28 && --& --& --& --\\[2mm]
\multirow{2}{*}{0.42}
& \multicolumn{1}{l}{\textsc{Gem++}}
 & 1.0 & 31 & 0.9 & 31 & -- & -- & -- & -- & -- & -- && 1.0 & 31 & 1.0 & 32 \\
& \multicolumn{1}{l}{\textsc{Gem90}}
 & 0.9 & 28 & 0.9 & 26 & -- & -- & -- & -- & -- & -- && --  & -- & --  & --\\

\end{tabular}
\end{ruledtabular}
  \renewcommand{\arraystretch}{1}
\caption{Diffusion plot: indicators $Q_R$ and $\Sigma_R$ of the
  global percent deviations between experimental data and calculations
  (with different parameters $y$, $t_{\mathrm{sw}}$, afterburner) for the
  mean and standard deviation, respectively,
  of the $R_v$ distribution in the considered range of $C_b$ (see text).
  The estimated statistical uncertainties are typically
  around 0.2$\%$ for $Q_R$ and 1--2$\%$ for $\Sigma_R$.
}
\label{tabdiff}
\end{table*}

At this point one can build a kind of diffusion plot for
binary events
%fulfilling the collinearity condition,
both for the experimental data and for the calculated results. 
The two-dimensional correlations of $(100~R_v)$ vs. $C_b$ produced by the
calculation are shown in Fig. \ref{diffusion}(a) and (d) for filtered events
from the reactions $^{93}$Nb+$^{93}$Nb and $^{93}$Nb+$^{116}$Sn, respectively.
A quantitative comparison with the experimental data of both systems
is shown in the remaining panels of Fig. \ref{diffusion}
for a few of the several calculations that were performed.
Panels (b) and (e) show, as a function of $C_b$, the average quantity
$\langle R_v \rangle$ and panels (c) and (f) the standard deviation
of the $R_v$ distribution.
Full (red) squares represent the experimental data, the other symbols
refer to calculations with $t_{\mathrm{sw}}$ = 500 fm/c, 
$y$ = 0.85 (0.42) and \textsc{Gem90} (\textsc{Gem++})
afterburner for the system Nb+Nb (Nb+Sn).
Circles are for filtered results, asterisks for unfiltered ones.

In the experimental data for $^{93}$Nb+$^{93}$Nb, $\langle R_v \rangle$
is substantially constant as a function of $C_b$ and close to 0.50
(as expected for a symmetric system), while in the asymmetric reaction
it is close to 0.445 (the projectile-to-total mass ratio).
Below $C_b \approx$ 1000 MeV the experimental $R_v$ shows a moderate
(and unexplained) increasing trend in $^{93}$Nb+$^{93}$Nb.
This effect is much more evident in the asymmetric $^{93}$Nb+$^{116}$Sn system.
However the calculation remains flat over the whole $C_b$ range in
$4\pi$ (asterisks) and, even after filtering (circles), it gives
a reasonably good reproduction of the data for $C_b >$ 1000 MeV.

The experimental standard deviation of the $R_v$ distribution
(rightmost panels) increases, as expected, with increasing centrality,
probably due to the growth of the fluctuations in the nucleon-nucleon
collision/exchange processes and in the secondary de-excitation.
The standard deviation for the calculated events reproduces the trend of the
experiments, although with a systematic underestimation.
It is worth noting that the value of the standard deviation is found to be
practically insensitive to $y$, $t_{\mathrm{sw}}$, afterburner and filtering.

More calculations with a wider choice of the parameters were performed,
but are not shown in Fig. \ref{diffusion} because their results do not
differ very much from the presented ones and would just blur the picture.

In order to summarize with a single number the quality of the agreement
between calculated results and experimental data for the mean value
of a generic observable $X$,
the average absolute percent deviation can be used,
\begin{equation}
  Q_X = \frac{100}{N}\; \sum_{i=1}^N
    \frac{\lvert \langle X\rangle _i^m - \langle X\rangle _i^e \rvert}
         {\langle X\rangle _i^e},
 \label{eq-Q}
\end{equation}
where $\langle X\rangle_i$ is the mean of the $X$ distribution for model
(apex $m$) and experiment (apex $e$) in the $i$-th bin of the sorting
variable (in our case $C_b$), and the summation index $i$ runs over the
chosen $N$ bins.
This indicator $Q_X$ measures the goodness of the global
agreement between experiment and calculations with different parameters:
the smaller the indicator, the better the agreement.
For the width of the $X$ distribution one can use a similar indicator
$\Sigma_X$, obtained from Eq. (\ref{eq-Q})
by replacing $\langle X\rangle _i^e$ and $\langle X\rangle _i^m$ with
$\langle \sigma(X) \rangle _i^e$ and $\langle \sigma(X) \rangle _i^m$,
respectively.

\begin{figure*}[htpb]        % FIG.4
\centering
\begin{tabular}{c}
\includegraphics[width=\textwidth]{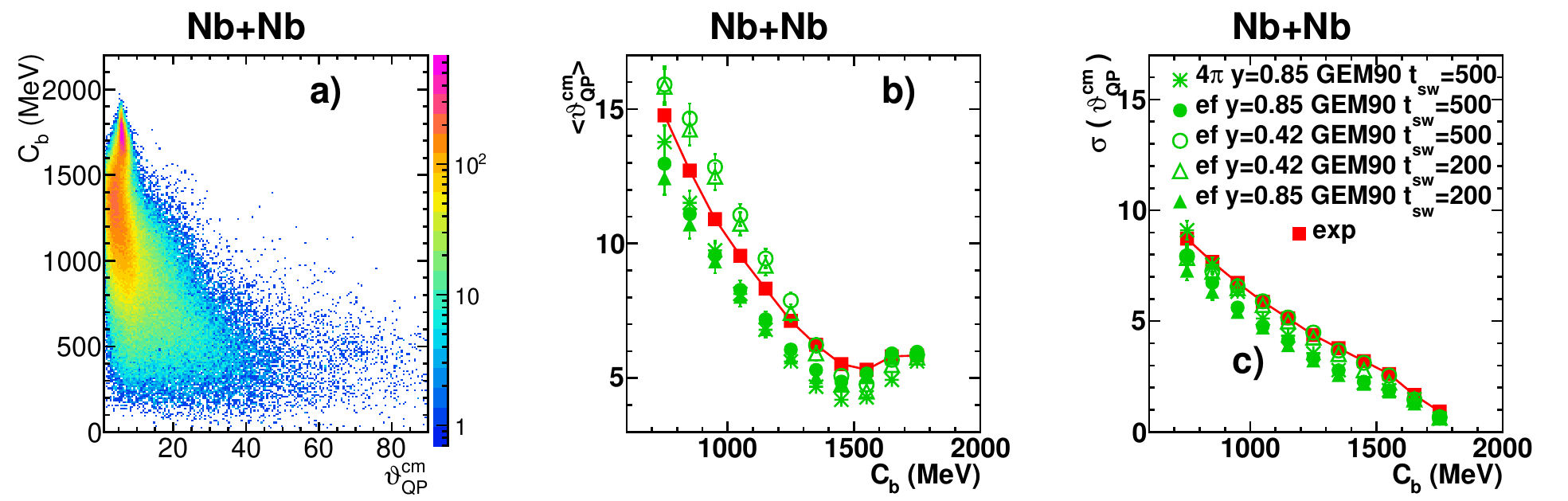}\\
\includegraphics[width=\textwidth]{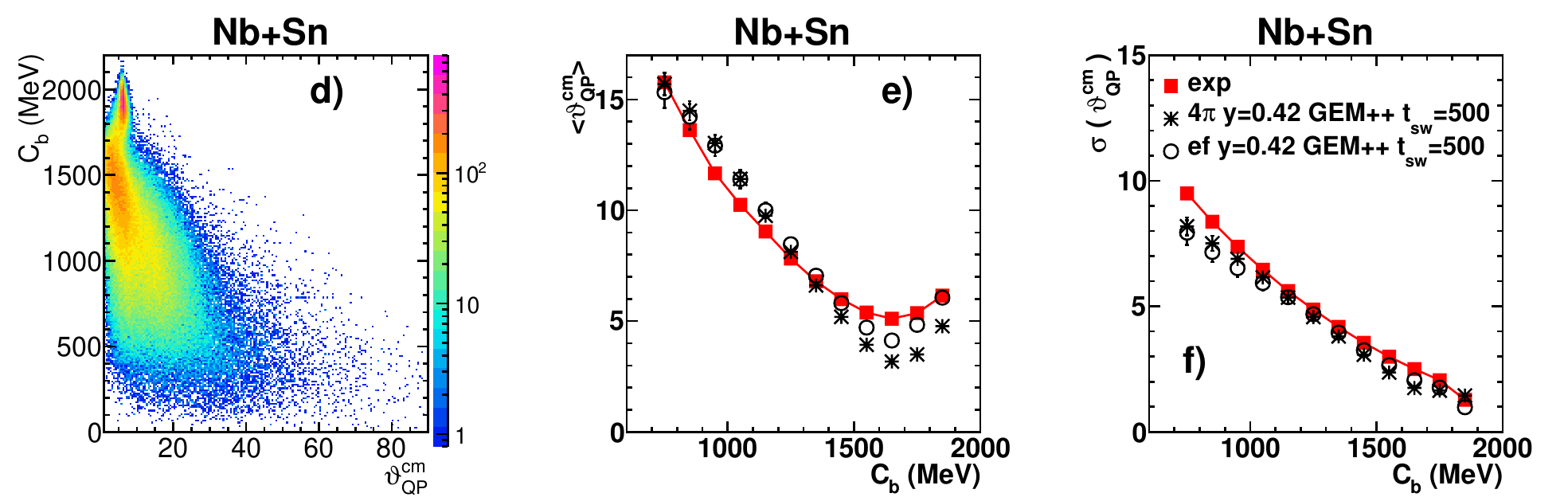}\\
\end{tabular}
\caption{(Color online) Wilczynski plot. Results of AMD plus \textsc{Gemini}
  calculations for the reactions $^{93}$Nb+$^{93}$Nb (upper panels) and
  $^{93}$Nb+$^{116}$Sn (lower panels) at 38 MeV/nucleon.
  (a) Correlation $\theta_{\mathrm{QP}}^{\,\mathrm{cm}}$ vs. $C_b$
   with parameters $y$=0.85, $t_{\mathrm{sw}}$=
   500 fm/c and \textsc{Gem++}; filtered results.
  (b) Mean of the $\theta_{\mathrm{QP}}^{\,\mathrm{cm}}$ distribution vs. $C_b$.
  (c) Standard deviation of the $\theta_{\mathrm{QP}}^{\,\mathrm{cm}}$
   distribution vs. $C_b$.
  (d) Same as panel (a) but for $^{93}$Nb+$^{116}$Sn with 
   parameters $y$=0.42, $t_{\mathrm{sw}}$= 500 fm/c and \textsc{Gem++};
   filtered results.
  (e) Same as panel (b) but for $^{93}$Nb+$^{116}$Sn.
  (f) Same as panel (c) but for $^{93}$Nb+$^{116}$Sn.
Full (red) squares are experimental data.
Other symbols for calculated results; they refer to filtered
events, except for asterisks, which are in $4 \pi$.}
\label{wilczynski}
\end{figure*}

\begin{table*}[htpb]    % Table II
  \begin{ruledtabular}
  \renewcommand{\arraystretch}{1.1}
\begin{tabular}{clccccccccccp{0.2mm}cccc}

\multicolumn{2}{c}{\multirow{2}{*}{}} &
\multicolumn{10}{c}{\bf Nb + Nb} &&\multicolumn{4}{c}{\bf Nb + Sn} \\ [1mm]

\multicolumn{1}{c}{} &
\multicolumn{1}{l}{$t_{\mathrm{sw}}$ (fm/c) =} &
\multicolumn{2}{c}{~~200~~} &
\multicolumn{2}{c}{~~500~~} &
\multicolumn{2}{c}{~1000~~} &
\multicolumn{2}{c}{~2500~~} &
\multicolumn{2}{c}{10000~~} & &
\multicolumn{2}{c}{~~200~~} &
\multicolumn{2}{c}{~~500~~} \\[1mm]

\cline{3-4} \cline{5-6} \cline{7-8} \cline{9-10} \cline{11-12}
                                   \cline{14-15} \cline{16-17}

\multicolumn{1}{c}{} &
\multicolumn{1}{c}{} &
\multicolumn{1}{c}{$Q_\theta$} &
\multicolumn{1}{c}{$\Sigma_\theta$} &
\multicolumn{1}{c}{$Q_\theta$} &
\multicolumn{1}{c}{$\Sigma_\theta$} &
\multicolumn{1}{c}{$Q_\theta$} &
\multicolumn{1}{c}{$\Sigma_\theta$} &
\multicolumn{1}{c}{$Q_\theta$} &
\multicolumn{1}{c}{$\Sigma_\theta$} &
\multicolumn{1}{c}{$Q_\theta$} &
\multicolumn{1}{c}{$\Sigma_\theta$} & &
\multicolumn{1}{c}{$Q_\theta$} &
\multicolumn{1}{c}{$\Sigma_\theta$} &
\multicolumn{1}{c}{$Q_\theta$} &
\multicolumn{1}{c}{$\Sigma_\theta$} \\

\multicolumn{1}{c}{$y$} &
\multicolumn{1}{c}{} &
\multicolumn{1}{c}{(\%)} &
\multicolumn{1}{c}{(\%)} &
\multicolumn{1}{c}{(\%)} &
\multicolumn{1}{c}{(\%)} &
\multicolumn{1}{c}{(\%)} &
\multicolumn{1}{c}{(\%)} &
\multicolumn{1}{c}{(\%)} &
\multicolumn{1}{c}{(\%)} &
\multicolumn{1}{c}{(\%)} &
\multicolumn{1}{c}{(\%)} & &
\multicolumn{1}{c}{(\%)} &
\multicolumn{1}{c}{(\%)} &
\multicolumn{1}{c}{(\%)} &
\multicolumn{1}{c}{(\%)} \\ [1mm]

\cline{1-17}\\ [-3mm]

\multirow{2}{*}{0.85}
& \multicolumn{1}{l}{\textsc{Gem++}}
 & 18 & 30 & 15 & 26 & 12 & 21 & 11 & 21 & 12 & 20 & & -- & -- & -- & -- \\
& \multicolumn{1}{l}{\textsc{Gem90}}
 & 14 & 24 & 11 & 21 & ~9 & 18 & ~9 & 18 & 10 & 19 & & -- & -- & -- & -- \\[2mm]
\multirow{2}{*}{0.42}
& \multicolumn{1}{l}{\textsc{Gem++}}
 & 10 & 15 & 10 & 11 & -- & -- & -- & -- & -- & -- & & ~8 & 14 & ~9 & 11\\
& \multicolumn{1}{l}{\textsc{Gem90}}
 & 10 & ~7 & 10 & ~4 & -- & -- & -- & -- & -- & -- & & -- & -- & -- & --\\

\end{tabular}
\end{ruledtabular}
\renewcommand{\arraystretch}{1}
\caption{Wilczynski plot: indicators $Q_\theta$ and $\Sigma_\theta$ of the
  global percent deviations between experimental data and calculations
  (with different parameters $y$, $t_{\mathrm{sw}}$, afterburner) for the
  mean and standard deviation, respectively,
  of the $\theta^{\,\mathrm{cm}}_{\mathrm{QP}}$ distribution
  in the considered range of $C_b$.
  The estimated statistical uncertainties are typically
  around 1--2$\%$ for both $Q_\theta$ and $\Sigma_\theta$.
 }
\label{tabwilc}
\end{table*}

\begin{figure*}[htpb]  % Fig. 5
\centering
\begin{tabular}{c}
\includegraphics[width=\textwidth]{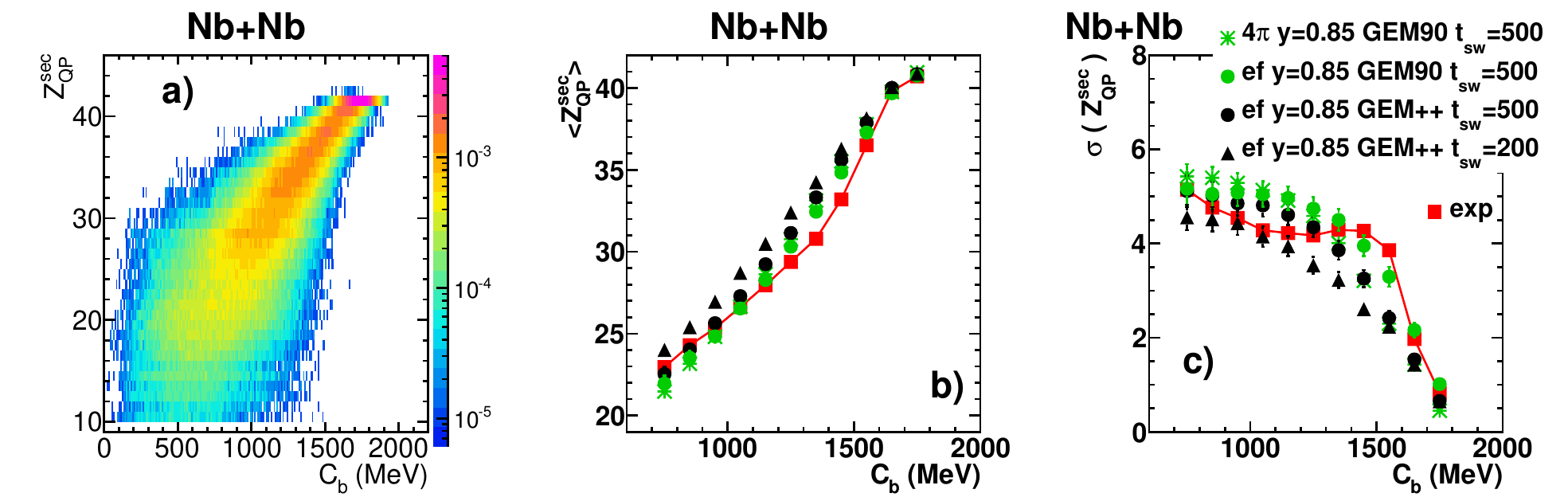}\\
\includegraphics[width=\textwidth]{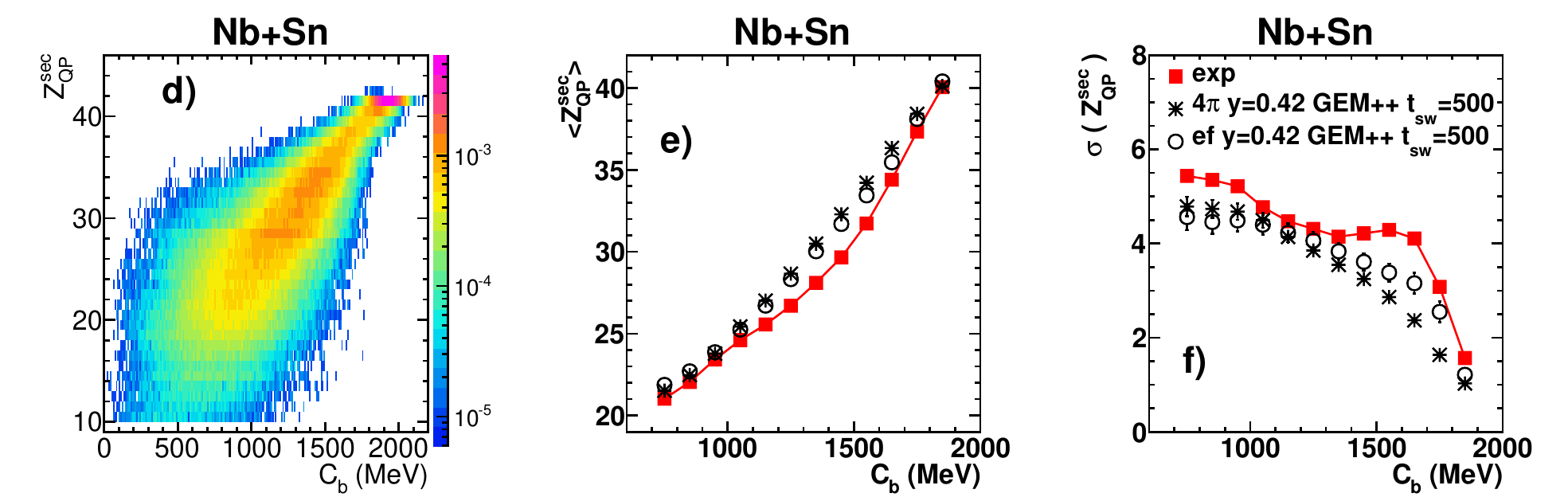}\\
\end{tabular}
\caption{(Color online) Secondary charge of QP.
  AMD plus \textsc{Gemini} calculations
  for the reactions $^{93}$Nb+$^{93}$Nb (upper panels) and
  $^{93}$Nb+$^{116}$Sn (lower panels) at 38 MeV/nucleon.
  (a) Correlation $Z_{QP}^{sec}$ vs. $C_b$, with
   $y$=0.85, $t_{\mathrm{sw}}$= 500 fm/c and \textsc{Gem++}; filtered results.
  (b) Mean of the $Z_{QP}^{sec}$ distribution vs. $C_b$.
  (c) Standard deviation of the $Z_{QP}^{sec}$ distribution vs. $C_b$.
  (d) Same as panel (a) but for the reaction $^{93}$Nb+$^{116}$Sn with
   $y$=0.42, $t_{\mathrm{sw}}$= 500 fm/c and \textsc{Gem++};
   filtered results.
  (e) Same as panel (b) but for the reaction $^{93}$Nb+$^{116}$Sn.
  (f) Same as panel (c) but for the reaction $^{93}$Nb+$^{116}$Sn.
Full (red) squares are experimental data of Ref. \cite{Piantelli2008}.
Other symbols are for calculated results; they refer to filtered
events, except for asterisks, which are in $4 \pi$.}
\label{zeta}
\end{figure*}

\begin{table*}[htpb]    % Table III
  \begin{ruledtabular}
  \renewcommand{\arraystretch}{1.1}
\begin{tabular}{clccccccccccp{0.2mm}cccc}

\multicolumn{2}{c}{\multirow{2}{*}{}} &
\multicolumn{10}{c}{\bf Nb + Nb} &&\multicolumn{4}{c}{\bf Nb + Sn} \\ [1mm]

\multicolumn{1}{c}{} &
\multicolumn{1}{l}{$t_{\mathrm{sw}}$ (fm/c) =} &
\multicolumn{2}{c}{~~200~~} &
\multicolumn{2}{c}{~~500~~} &
\multicolumn{2}{c}{~1000~~} &
\multicolumn{2}{c}{~2500~~} &
\multicolumn{2}{c}{10000~~} & &
\multicolumn{2}{c}{~~200~~} &
\multicolumn{2}{c}{~~500~~} \\[1mm]

\cline{3-4} \cline{5-6} \cline{7-8} \cline{9-10} \cline{11-12}
                                   \cline{14-15} \cline{16-17}

\multicolumn{1}{c}{} &
\multicolumn{1}{c}{} &
\multicolumn{1}{c}{$Q_Z$} &
\multicolumn{1}{c}{$\Sigma_Z$} &
\multicolumn{1}{c}{$Q_Z$} &
\multicolumn{1}{c}{$\Sigma_Z$} &
\multicolumn{1}{c}{$Q_Z$} &
\multicolumn{1}{c}{$\Sigma_Z$} &
\multicolumn{1}{c}{$Q_Z$} &
\multicolumn{1}{c}{$\Sigma_Z$} &
\multicolumn{1}{c}{$Q_Z$} &
\multicolumn{1}{c}{$\Sigma_Z$} & &
\multicolumn{1}{c}{$Q_Z$} &
\multicolumn{1}{c}{$\Sigma_Z$} &
\multicolumn{1}{c}{$Q_Z$} &
\multicolumn{1}{c}{$\Sigma_Z$} \\

\multicolumn{1}{c}{$y$} &
\multicolumn{1}{c}{} &
\multicolumn{1}{c}{(\%)} &
\multicolumn{1}{c}{(\%)} &
\multicolumn{1}{c}{(\%)} &
\multicolumn{1}{c}{(\%)} &
\multicolumn{1}{c}{(\%)} &
\multicolumn{1}{c}{(\%)} &
\multicolumn{1}{c}{(\%)} &
\multicolumn{1}{c}{(\%)} &
\multicolumn{1}{c}{(\%)} &
\multicolumn{1}{c}{(\%)} & &
\multicolumn{1}{c}{(\%)} &
\multicolumn{1}{c}{(\%)} &
\multicolumn{1}{c}{(\%)} &
\multicolumn{1}{c}{(\%)} \\ [1mm]

\cline{1-17}\\ [-3mm]

\multirow{2}{*}{0.85}
& \multicolumn{1}{l}{\textsc{Gem++}}
 & 6.8 & 18 & 3.7 & 13 & 3.3 & 12 & 3.2 & 11 & 3.5 & 14 &
 & --  & -- & --  & -- \\
& \multicolumn{1}{l}{\textsc{Gem90}}
 & 4.3 & ~9 & 2.7 & 10 & 2.9 & 12 & 2.7 & 12 & 2.6 & 11 &
 & --  & -- & --  & -- \\ [2mm]
\multirow{2}{*}{0.42}
& \multicolumn{1}{l}{\textsc{Gem++}}
 & 6.5 & 14 & 4.6 & 12 & --  & -- & --  & -- &  --  & -- &
 & 7.1 & 17 & 4.2 & 14 \\
& \multicolumn{1}{l}{\textsc{Gem90}}
 & 3.6 & ~9 & 1.9 & ~8 & --  & -- & --  & -- &  --  & -- &
 & --  & -- & --  &  --  \\

\end{tabular}
\end{ruledtabular}
\renewcommand{\arraystretch}{1}
\caption{$Z_{QP}$ vs. $C_b$ correlation: indicators $Q_Z$ and $\Sigma_Z$ of
  the global percent deviations between experimental data and calculations
  (with different parameters $y$, $t_{\mathrm{sw}}$, afterburner) for the
  mean and standard deviation, respectively,
  of the $Z_{\mathrm{QP}}$ distribution in the considered range of $C_b$.
  Estimated statistical uncertainties are typically
  around 0.4$\%$ for $Q_Z$ and 2$\%$ for $\Sigma_Z$.
}
\label{tabzsec}
\end{table*}

The quality of the agreement between the experimental diffusion plot
and those obtained with {\em all } the different calculations that were
performed (varying $y$, $t_{\mathrm{sw}}$ and afterburner) is presented in
Table \ref{tabdiff} \footnote{Here and in the following tables, the point
  with the largest $C_b$ value is not used for calculating the indicators,
  because in the experiment it is strongly polluted by elastic events;
  the considered $C_b$ range will be 700--1700
  (700--1800) MeV for Nb+Nb (Nb+Sn).},
by means of the above defined indicators,
where the observable $X$ is now $R_v$.
A quick glance at the table shows that:
  i) the difference in $\langle R_v \rangle$ between experiment and
  calculations is always very small, around  0.8--1.0$\%$
  (even in the asymmetric system where this outcome is not trivial);
  ii) the calculations underestimate the standard deviations $\sigma(R_v)$
  by about  30$\%$;
  iii) for what concerns the observable $R_v$, the calculations with
  different parameters do not substantially differ from each other.

The two-dimensional correlations between $\theta_{\mathrm{QP}}^{\,\mathrm{cm}}$
and $C_b$ (Wilczynski plot) produced by the calculations are shown in
Fig. \ref{wilczynski}(a) and (d) for filtered events from $^{93}$Nb+$^{93}$Nb
and $^{93}$Nb+$^{116}$Sn, respectively, with the same parameters used for
Fig. \ref{diffusion}(a) and (d).
In the remaining panels of Fig. \ref{wilczynski}, the full (red) squares
show, again as a function of $C_b$, the mean [(b) and (e)] and the
standard deviation [(c) and (f)] of the $\theta_{\mathrm{QP}}^{\,\mathrm{cm}}$
distributions for the two reactions. 
One observes the typical behavior of binary processes at intermediate
energies: with decreasing $C_b$ (i.e. with increasing centrality) the
experimental average value of the polar angle first starts close to the
grazing angle of very peripheral collisions, then approaches the
beam axis and finally moves definitively away from it;
at the same time the standard deviation increases monotonically,
as expected on the basis of the growing importance of the fluctuations
induced by nucleon-nucleon collisions and secondary decays.
The other symbols are the results of some (out of several) calculations
that were performed with different combinations of the parameters
$y$, $t_{\mathrm{sw}}$ and afterburner.
The effect of the experimental filter is again negligible.
The calculations do a good job by reproducing very closely the
behavior of the experimental data.

Focusing on the symmetric collision $^{93}$Nb+$^{93}$Nb, where calculations
with two values of $y$ were performed, one observes some sensitivity
to a variation of the in-medium nucleon-nucleon cross section, although
the results for the mean value of $\theta_{\mathrm{QP}}^{\,\mathrm{cm}}$
do not seem conclusive.
A small improvement with $y$ = 0.42 is observed for the standard deviation,
but its interpretation is not clear.
In any case, it will be shown in Subsect. \ref{lcp} that 
$y$ = 0.42 gives a slightly worse reproduction of the chemistry
of emitted LCPs and IMFs.
  
The obtained results for all performed calculations are 
summarized in Table \ref{tabwilc}, where the deviations between
experimental data and different simulations are shown in terms of the
already defined indicators, evaluated for the observable
$X \equiv \theta^{\mathrm{cm}}_{\mathrm{QP}}$.
For the case $y$= 0.85 and with both \textsc{Gemini} versions, one
observes that the agreement with the experimental data
first improves with increasing $t_{\mathrm{sw}}$ and then stabilizes.
For example, with \textsc{Gem++}, $Q_\theta$ starts with 18\% at
$t_{\mathrm{sw}}$ = 200 fm/c and then decreases to 15\% and $\sim$12\%
for longer times.
This is attributed to the already mentioned incomplete Coulomb deflection
when the switching from AMD to \textsc{Gemini} occurs too early.
Using \textsc{Gem90} brings a general, limited improvement, while the
trend as a function of switching time remains the same.
Also the $\Sigma_\theta$ indicator for the
%differences in the
standard deviations displays the same behavior (slightly lower values
with \textsc{Gem90} and some improvement with larger $t_{\mathrm{sw}}$),
although its interpretation is not as obvious.

A further observable that can be investigated, although in a limited
range of polar angles, is the charge $Z_{QP}^{sec}$ of the QP, detected
by means of the Silicon telescopes.
The calculated two-dimensional correlations between $Z_{QP}^{sec}$ and $C_b$
are shown in Fig. \ref{zeta}(a) and (d) for the reactions
$^{93}$Nb+$^{93}$Nb and $^{93}$Nb+$^{116}$Sn, respectively,
with the same parameters used for Fig. \ref{diffusion}(a) and (d).
In the remaining panels of Fig. \ref{zeta} the full (red) squares show,
as a function of $C_b$, the mean [(b) and (e)] and the standard
deviation [(c) and (f)] of the distribution of
the experimental charge $Z_{QP}^{sec}$.
\begin{figure*}[htpb]     % FIG.6
\centering
\begin{tabular}{c}
\includegraphics[width=\textwidth]{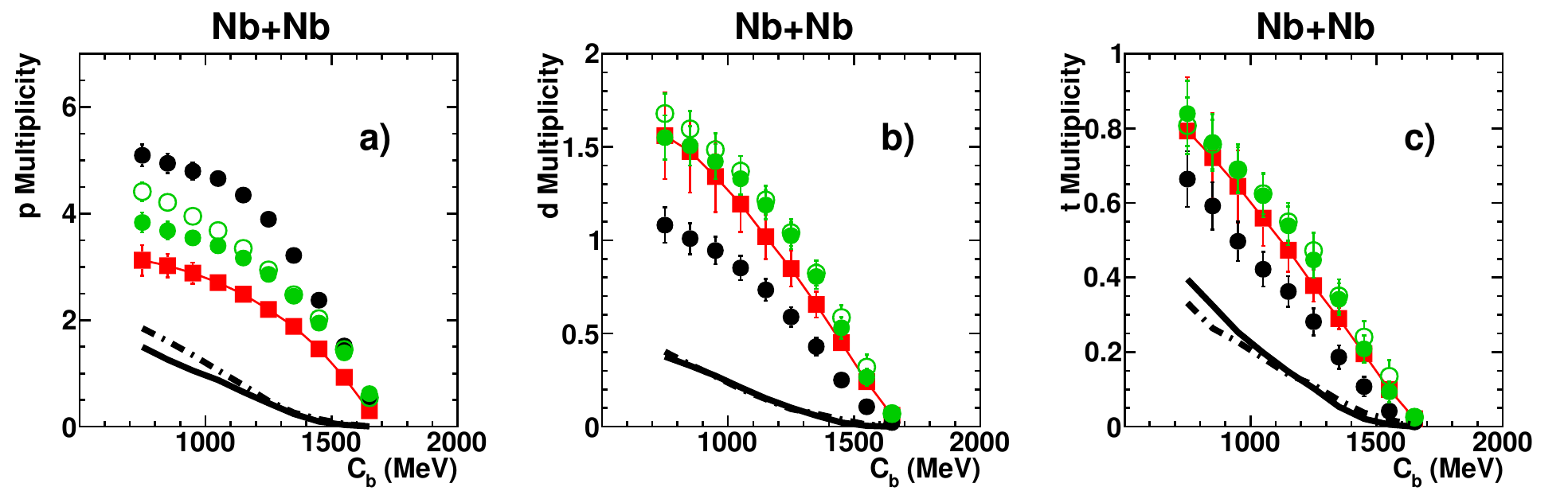}\\
\includegraphics[width=0.7\textwidth]{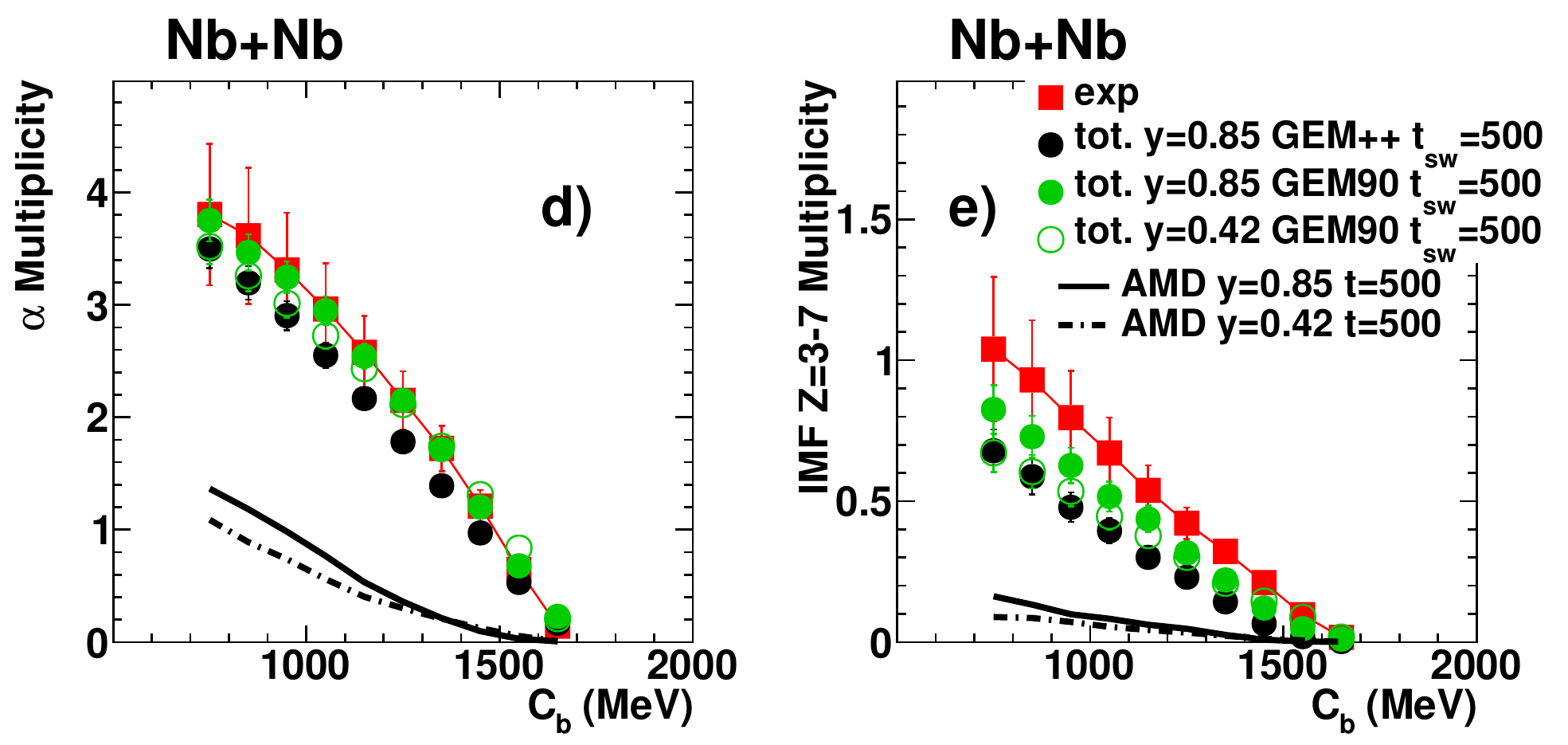}\\
\end{tabular}
\caption{(Color online) Multiplicities of LCPs and IMFs ($Z$=3--7),
  forward-emitted in the c.m. frame, as a function of $C_b$ for the
  system $^{93}$Nb+$^{93}$Nb at 38 MeV/nucleon:
  (a) protons, (b) deuterons, (c) tritons,
  (d) $\alpha$ particles and (e) IMF ($Z$=3--7).
  Experimental data of Ref. \cite{Piantelli2006} [full (red) squares] are
  compared with the results of calculations, all with $t_{\mathrm{sw}}$= 500
  fm/c, but different $y$ (0.85 or 0.42) or afterburner (\textsc{Gem90}
  or \textsc{Gem++}).
  Circles correspond to AMD plus \textsc{Gemini}
  (\textsc{Gem++} in black, \textsc{Gem90} in green),
  lines to results of AMD alone, stopped at the indicated time $t$.
}
\label{molt1}
\end{figure*}
The calculated results are shown with different symbols corresponding
to different parametrizations.
As expected, with increasing centrality of the collision,
%(i.e. decreasing $C_b$),
the mean of the $Z_{QP}^{sec}$ distribution steadily
decreases, starting from the charge $Z$ = 41 of the projectile in the
most peripheral collisions, while the standard deviation increases.
Figure \ref{zeta} indicates that the agreement between experimental data
and calculation is rather good for the means, while some
systematic difference is present in the standard deviations.

As for the previous variables, Table \ref{tabzsec} summarizes the deviation
between the experimental data and all the calculated results by means of
the quality indicators now applied to the observable $X \equiv Z_{QP}^{sec}$.
There is a clear improvement when using \textsc{Gem90} instead
of \textsc{Gem++}.
For example, for the calculation with $y$ = 0.85 and $t_{\mathrm{sw}}$ =
200 fm/c, the indicators
$Q_Z$ and $\Sigma_Z$ decrease by about a factor of two.
Focusing on the case $y$ = 0.85 and \textsc{Gem90}, one observes
an improvement when $t_{\mathrm{sw}}$ is increased from 200 to 500 fm/c,
but longer times do not bring any further change.
This suggests that the emissions of AMD between 200 and 500
fm/c differ from the emissions of \textsc{Gemini},
but this difference becomes negligible beyond 500 fm/c.
The charge of the QP shows little sensitivity to a reduction of
the in-medium cross section from $y$= 0.85 to 0.42.
However, with $y$ = 0.42 the difference between 200 and 500 fm/c
is reduced with respect to what happens with $y$ = 0.85.
This may be explained by the fact that lowering the in-medium cross
section produces less nucleon-nucleon collisions and therefore
attenuates the difference between dynamic and statistical emissions
in the time interval between 200 and 500 fm/c.

%--------------------------------------------------------------------

\subsection{Particle multiplicities}
\label{lcp}

\begin{figure*}[htpb]     % FIG.7
\centering
\includegraphics[width=\textwidth]{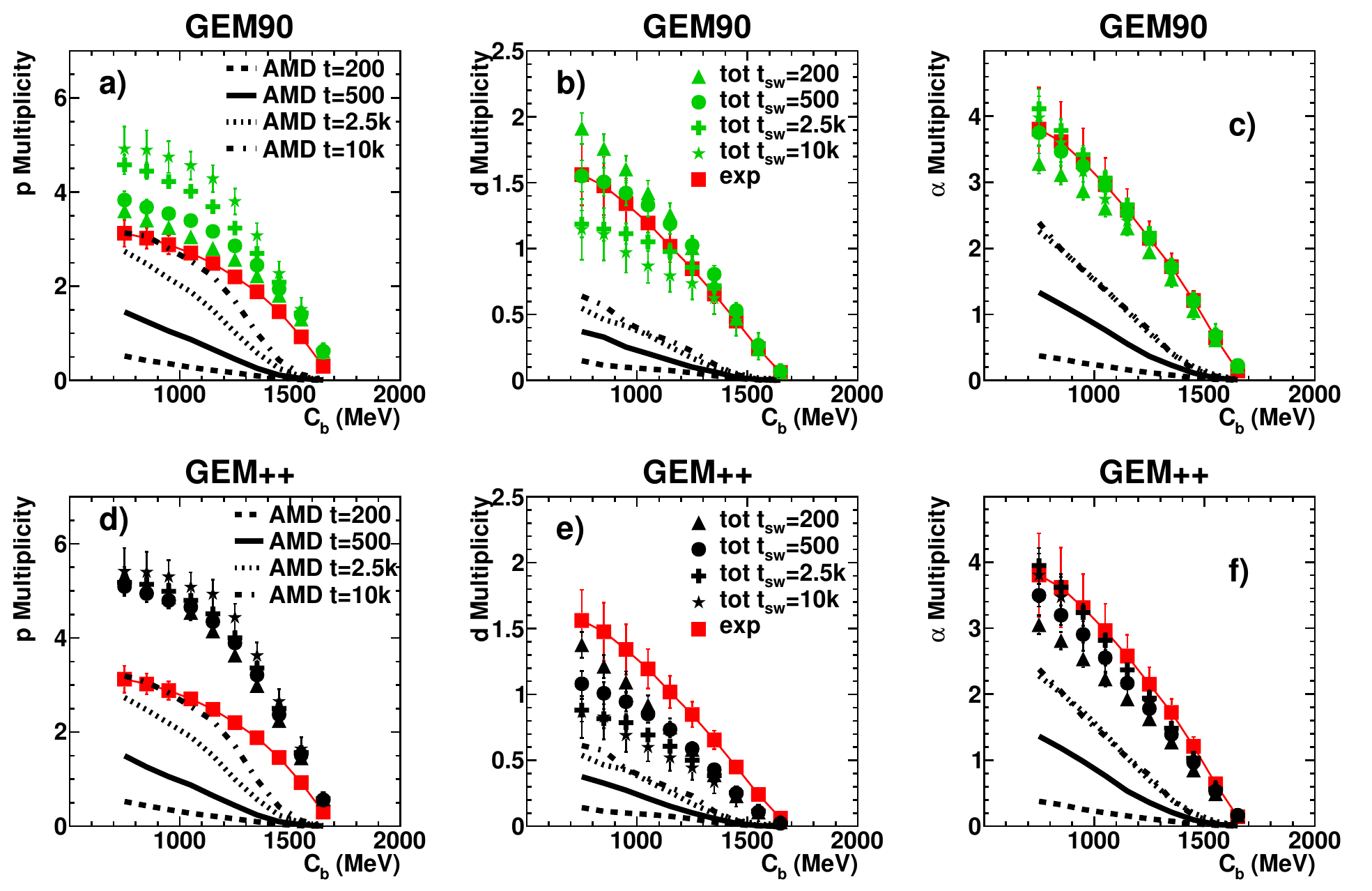}
\caption{(Color online) Multiplicities of LCPs,
  forward-emitted in the c.m. frame, as a function of $C_b$ for the system
  $^{93}$Nb+$^{93}$Nb at 38 MeV/nucleon:
  (a), (d) protons; (b), (e) deuterons; (c), (f) $\alpha$ particles.
  Experimental data [full (red) squares] are compared with results of
  calculations with $y$=0.85 and different values of $t_{\mathrm{sw}}$
  (in fm/c).
  Symbols correspond to AMD plus \textsc{Gemini}
  (\textsc{Gem90} in upper panels, \textsc{Gem++} in lower ones),
  black lines to results of AMD alone, stopped at the indicated time $t$.
  }
\label{molt2}
\end{figure*}

As a last point, the experimental total multiplicities of LCP ($Z$ = 1, 2)
and IMF ($Z$ = 3--7) of Ref. \cite{Piantelli2006},
associated with 2-body events from the reaction $^{93}$Nb+$^{93}$Nb,
have been used for comparison with the total multiplicities obtained from
AMD plus \textsc{Gemini} calculations
\footnote{The separation between midvelocity and evaporative multiplicities
  is beyond the scope of this paper and will not be discussed.}.

Figure \ref{molt1} displays the multiplicity of particles emitted in the
forward c.m. hemisphere as a function of $C_b$, separately for the
different species: protons in panel (a), deuterons in (b),
tritons in (c), $\alpha$ particles in (d), IMF with $Z$ = 3--7 in (e).
Both the experimental data [full (red) squares] from Fig. 3
of \cite{Piantelli2006}
and the calculated results with $t_{\mathrm{sw}}$= 500 fm/c are presented.
The circles represent the total multiplicities
obtained by taking into account the emissions of the \textsc{Gemini}
afterburner, while the lines show the particles multiplicities
produced by the dynamic code AMD, stopped at the indicated time $t$.
With the exception of the protons (which are always overestimated by the
calculations) and to a minor extent of the IMFs,
the remaining experimental multiplicities are quite well reproduced by the
AMD model with $y$=0.85, coupled with \textsc{Gem90} [full (green) circles].
In fact, with the \textsc{Gem90} afterburner one observes a clear improvement
with respect to \textsc{Gem++} (full black circles) even for protons, because
their calculated multiplicity is strongly reduced and approaches the
experimental data from above.
A relevant difference between \textsc{Gem90} and \textsc{Gem++} lies in the
adopted expression for the
level density, which is that of Ref. \cite{Charity10} for
\textsc{Gem++} and that of Ref. \cite{Fineman91} for \textsc{Gem90}.
In fact, it was verified that if the level density parametrization of
\textsc{Gem90} is implemented in \textsc{Gem++}, the difference
between the results of the two versions is strongly reduced.
A possible explanation of the better agreement with \textsc{Gem90} 
may be due to the fact that \textsc{Gem++} was optimized
so as to reproduce experimental data for rather heavy nuclei
(mainly in the range $A \approx$ 100--200 \cite{Charity10}),
while the decaying QP of Ref. \cite{Piantelli2006} are likely to have
masses in the range $A \approx$ 40--93.

\begin{table}[htpb]    % Table IV
  \begin{ruledtabular}
  \renewcommand{\arraystretch}{1.20}

\begin{tabular}{lclccccc}
\multicolumn{2}{c}{\multirow{2}{*}{}} &
\multicolumn{1}{r}{$t_{\mathrm{sw}}$ (fm/c) :} &
\multicolumn{1}{c}{~~200} &
\multicolumn{2}{c}{~~500} &
\multicolumn{1}{c}{~2500} &
\multicolumn{1}{c}{10000} \\

\cline{5-6}
\multicolumn{2}{c}{} &
\multicolumn{1}{r}{asy-EoS :} &
\multicolumn{1}{c}{stiff} &
\multicolumn{1}{c}{stiff} &
\multicolumn{1}{c}{soft~} &
\multicolumn{1}{c}{stiff} &
\multicolumn{1}{c}{stiff~~} \\ [2mm]

\multicolumn{1}{l}{} &
\multicolumn{1}{c}{} &
\multicolumn{1}{c}{} &
\multicolumn{1}{c}{$Q_M$} &
\multicolumn{1}{c}{$Q_M$} &
\multicolumn{1}{c}{$Q_M$} &
\multicolumn{1}{c}{$Q_M$} &
\multicolumn{1}{c}{$Q_M$} \\

\multicolumn{1}{l}{} &
\multicolumn{1}{c}{$y$} &
\multicolumn{1}{c}{} &
\multicolumn{1}{c}{(\%)} &
\multicolumn{1}{c}{(\%)} &
\multicolumn{1}{c}{(\%)} &
\multicolumn{1}{c}{(\%)} &
\multicolumn{1}{c}{(\%)} \\

\hline
\multirow{4}{*}{p}
%                    200   500stif  500soft  2500  10000
  & \multirow{2}{*}{0.85}
  & \textsc{Gem++} & 64 & 68 & 64 & 74 & 86 \\
 && \textsc{Gem90} & 18 & 29 & 24 & 47 & 64 \\ [2mm]
  & \multirow{2}{*}{0.42}
  & \textsc{Gem++} & 64 & 80 & -- & -- & -- \\
 && \textsc{Gem90} & 20 & 39 & -- & -- & -- \\
\hline
\multirow{4}{*}{d}
  & \multirow{2}{*}{0.85}
  & \textsc{Gem++} & 30 & 35 & 19 & 43 & 49 \\
 && \textsc{Gem90} & 16 & 12 & 28 & 13 & 17 \\ [2mm]
  & \multirow{2}{*}{0.42}
  & \textsc{Gem++} & 30 & 31 & -- & -- & -- \\
 && \textsc{Gem90} & 16 & 19 & -- & -- & -- \\
\hline
\multirow{4}{*}{t}
  & \multirow{2}{*}{0.85}
  & \textsc{Gem++} & 33 & 30 & 23 & 27 & 28 \\
 && \textsc{Gem90} & ~8 & 10 & 16 & ~6 & 12 \\ [2mm]
  & \multirow{2}{*}{0.42}
  & \textsc{Gem++} & 29 & 25 & -- & -- & -- \\
 && \textsc{Gem90} & 11 & 16 & -- & -- & -- \\
\hline
\multirow{4}{*}{$\alpha$}
  & \multirow{2}{*}{0.85}
  & \textsc{Gem++} & 24 & 15 & 22 & ~8 & 13 \\
 && \textsc{Gem90} & 11 & ~2 & ~6 & ~4 & ~5 \\ [2mm]
  & \multirow{2}{*}{0.42}
  & \textsc{Gem++} & 23 & 17 & -- & -- & -- \\
 && \textsc{Gem90} & 11 & ~9 & -- & -- & -- \\
\hline
\multirow{4}{*}{IMFs}
  & \multirow{2}{*}{0.85}
  & \textsc{Gem++} & 62 & 50 & 53 & 53 & 50 \\
 && \textsc{Gem90} & 40 & 28 & 33 & 42 & 44 \\ [2mm]
  & \multirow{2}{*}{0.42}
  & \textsc{Gem++} & 60 & 56 & -- & -- & -- \\
 && \textsc{Gem90} & 32 & 30 & -- & -- & -- \\

\end{tabular}
\end{ruledtabular}
\renewcommand{\arraystretch}{1}
\caption{Multiplicities of LCPs and IMFs ($Z$ = 3--7) in the reaction
    $^{93}$Nb + $^{93}$Nb at 38 MeV/nucleon: indicators $Q_M$ of the
    average percent deviation between experimental data of
    Ref. \cite{Piantelli2006} and calculated multiplicities
    (with different parameters $y$, $t_{\mathrm{sw}}$, afterburner
    and stiff/soft equation of state)
    in the considered range of $C_b$.
    Typical estimated statistical uncertainties on $Q_M$ are
    of the order of 4--8$\%$.}
\label{tabmult}
\end{table}

As mentioned in subsection \ref{QP}, the decrease of $y$ from 0.85
(full green circles) to 0.42 (open green circles) tends to slightly worsen
the quality of the agreement between calculated results and experimental data, 
mainly for less peripheral collisions and especially for protons and
$\alpha$ particles.
Another point worth noting is that
a sizable amount of particles is directly produced by AMD (lines).
Of course, such a contribution, which is produced with a dynamic formalism,
originates partly during the interaction of the colliding nuclei,
but partly also after their reseparation.

The evolution of the multiplicities for protons, deuterons and
$\alpha$ particles when the switching times increase from 200 up to
10000 fm/c is presented in Fig. \ref{molt2}.
The upper panels refer to \textsc{Gem90}, the lower ones to \textsc{Gem++}.
Again, the lines are the multiplicities from AMD without afterburner,
obtained with the standard value $y$=0.85 for the in-medium cross section.
They show a general increase from 200 fm/c (dashed lines) to 500 fm/c
(full lines) and 2500 fm/c (dotted lines), while a further delay of
$t_{\mathrm{sw}}$ up to 10000 fm/c (dash-dotted lines) does not produce any
appreciable effect, with the notable exception of protons.
Another point worth noting is that at $t_{\mathrm{sw}}$= 10000 fm/c,
the protons produced by the AMD alone in the less peripheral collisions
already reach the experimental data, therefore the addition of the
evaporative emissions with whatever \textsc{Gemini} version leads to an
overestimation of the data.
Thus, at very long times, AMD seems to have some problems,
which are put in particular evidence by the very large production of protons.

\begin{figure}[bhtp]   % FIG.8
\centering
%\begin{tabular}{c}
%\includegraphics[width=0.48\textwidth]{ztotcpp.pdf} \\
%\includegraphics[width=0.48\textwidth]{ztotf90.pdf}
\includegraphics[width=0.48\textwidth]{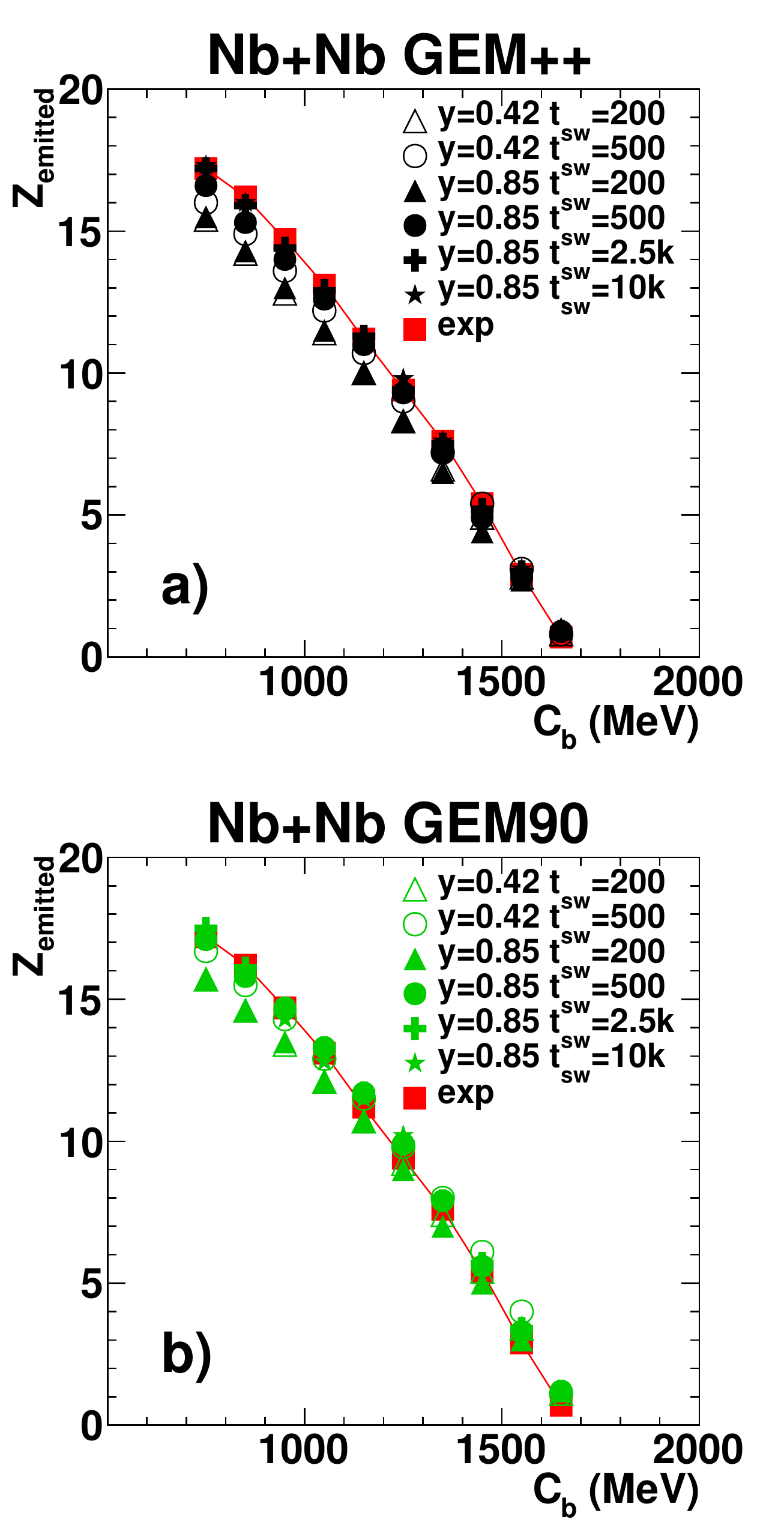}
%\includegraphics[width=\textwidth]{ztot4.pdf}
%\end{tabular}
\caption{(Color online) Total charge of all LCPs and IMFs ($Z$=3--7)
  forward-emitted in the c.m. reference frame of the collision
  for the $^{93}$Nb+$^{93}$Nb reaction.
  Full (red) squares represent experimental data, the other symbols
  results of various calculations
  with \textsc{Gem++} (upper) and \textsc{Gem90} (lower) afterburner.
  Switching times are in units of fm/c.}
\label{zetatot}
\end{figure}

With AMD plus \textsc{Gem90} (upper panels),
the total multiplicities of $\alpha$ particles and IMFs (not shown) are
almost independent of switching times, as one would expect for a good
matching of the two models.
However this does not happen for protons and deuterons, which display
opposite trends with switching time:
when $t_{\mathrm{sw}}$ increases, the total multiplicity of protons
increases too (reaching an overestimation of the experimental data
of about 50$\%$), while that of deuterons decreases (leading to an
underestimation of about 50$\%$ at long times).
This means that in the time interval between 200 and 10000 fm/c
AMD tends to produce more protons (and less
deuterons) than \textsc{Gem90}, while the two codes
produce similar amounts of more complex particles.

\begin{figure*}[bhpt]    % FIG.9
\centering
\includegraphics[width=\textwidth]{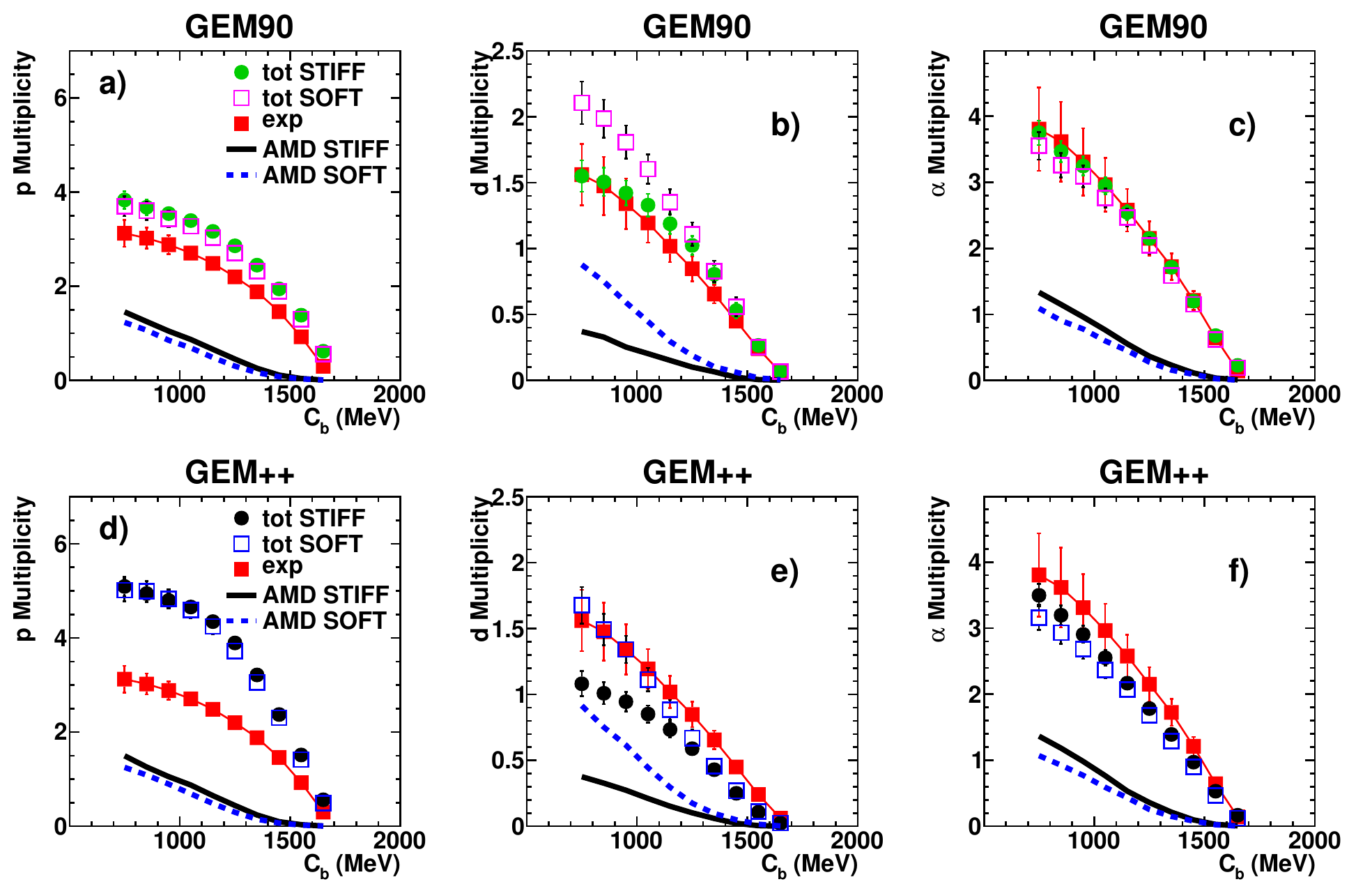}
\caption{(Color online) Same as Fig. \ref{molt2}, but here experimental
  data [full (red) squares] are compared with calculations 
  ($y$=0.85 and $t_{\mathrm{sw}}$= 500 fm/c) performed with an asy-stiff
  and an asy-soft version of the equation of state in the AMD model.
  Lines (symbols) correspond to results of AMD without (with) afterburner.
  The asy-stiff results are those of Fig. \ref{molt2},
  presented with the same lines/symbols and colors.
  Upper panels refer to \textsc{Gem90}, lower panels to \textsc{Gem++}.}
\label{molt3}
\end{figure*}

The total multiplicities of the various particles obtained with
AMD plus \textsc{Gem++}
(symbols in the lower panels) appear to be much less dependent on
switching times:
the case of protons, where all symbols are practically superposed, is
especially striking and also the results for the deuterons are less
spread out than with \textsc{Gem90}.
This may be an indication that the particle production of AMD at long times
resembles more the evaporation described by \textsc{Gem++},
than by \textsc{Gem90}.
However the results with \textsc{Gem++} present a much worse agreement
with the data: at all switching times the multiplicity of deuterons
underestimates the experimental data and that of the protons overestimates
them by a good factor of two.

All results concerning the global deviations
(indicator $Q_M$) of the calculated particle multiplicities from the
experimental values of Ref. \cite{Piantelli2006} are presented in
Table \ref{tabmult} for different values of the calculation parameters.
The entries of this table confirm that using \textsc{Gem90},
the deviations from the experimental values are sizably reduced
for all particles and practically at all switching times.
For tritons and $\alpha$ particles $Q_M$ becomes comparable
or even smaller than its estimated uncertainty, thus indicating a very
good agreement between model and experiment, within errors.
Thus, for what concerns the particle multiplicities, the calculation
that best reproduces all experimental data seems to be AMD coupled with
the \textsc{Gem90} afterburner, with a switching time of about 500 fm/c
and a preference for $y$ = 0.85.

Figure \ref{zetatot} presents the total charge of LCPs plus IMFs
($Z$ = 3--7), forward-emitted in the c.m. frame
of the colliding system, obtained by adding the charge of the
particles weighted with their respective multiplicities.
First of all, it is worth noting the good agreement,
for almost all choices of parameters,
of the calculated total charge $Z_{\mathrm{emitted}}$ with the measured one,
which is again represented by the full (red) squares that are hardly
visible below the symbols of the various calculations.
The two versions of \textsc{Gemini} [GEM++ in (a) and GEM90 in (b)] reproduce
equally well the total emitted charge, although they give a
different chemistry of the emitted particles (see Fig. \ref{molt1}).
Second, the calculated $Z_{\mathrm{emitted}}$ is
practically independent of the switching time of the dynamic calculation,
with the possible exception of $t_{\mathrm{sw}}$=200 fm/c (triangles),
for which the total emitted charge is somewhat underestimated.
As already noted about Table \ref{tabzsec},
this may be an indication of dynamic effects between 200 and 500 fm/c.
In fact the calculations somewhat overestimate the final charge
$Z_{QP}^{sec}$ of the QP and underestimate by the same amount the
total charge of LCPs and IMFs in such a way that they still sum up
to $Z \approx$ 41.
Third, the use of $y$=0.42 slightly reduces the total emitted charge
(open symbols compared with full symbols)
in the less peripheral collisions, and this again may be attributed to
a reduced importance of dynamic effects at short times and
to lower dissipation of kinetic energy, resulting in less excited nuclei.

A last point to be addressed is the isospin dependence of the equation of state.
The isospin degree of freedom plays an important role in determining
the exit channels.
In fact isospin transport phenomena, related to the isospin gradient
between target and projectile (isospin diffusion
\cite{DiToro2010,Baran05}) or to the density gradient between QP--QT
(which are both at normal density) and the more diluted midvelocity zone
(isospin drift \cite{Baran05}) have been observed in several experiments
(see, e.g., Refs. \cite{Tsang2004,Tsang2009,Lombardo2010,DeFilippo2012,Barlini2013,Piantelli2017,Jedele2017,Manso2017}).
Of course, some sensitivity to the effects of an asy-stiff or asy-soft
interaction in the AMD calculations may be expected only for mass-resolved
light particles.
Figure \ref{molt3} shows the results with AMD only (lines) or with AMD plus
\textsc{Gemini} afterburner (symbols), for the two hypotheses about
the equation of state mentioned at the beginning of Sect. \ref{AMD-model}.
The panels show again results for protons [(a) and (d)],
deuterons [(b) and (e)] and $\alpha$ particles [(c) and (f)],
with \textsc{Gem90} (upper panels) and \textsc{Gem++} (lower panels),
and experimental values again represented by full (red) squares.

In the dynamic calculations without afterburner, protons and
$\alpha$ particles present a very small difference between the
asy-stiff (full lines) and asy-soft (dashed lines) interaction,
and this insensitivity remains also after the Gemini decay.
Apparently, there is a remarkable signal for deuterons,
which are produced more abundantly (by a factor of about two) with
the asy-soft equation of state, as found also in Ref. \cite{OnoJPC2013}.
Actually this effect, which at first sight might seem to favor an
asy-soft AMD plus GEM++, is very likely to be an artifact.
Its explanation resides in
the fact that when the soft SLy4 force \cite{ChabanatSLy4} is
used in the present version of AMD, one finds that
it largely overestimates the deuteron binding energy.
On the contrary, the stiff parametrization gives a deuteron binding energy
much closer to the true value and it also reproduces rather well
the measured deuteron multiplicities with GEM90.

%--------------------------------------------------------------------

\section{Summary and conclusions}

In this work the properties of quasi-projectiles (QP) detected in
peripheral and semiperipheral collisions of the reactions
$^{93}$Nb+$^{93}$Nb and $^{93}$Nb+$^{116}$Sn at 38 MeV/nucleon
have been compared with calculations performed with the dynamic code AMD,
followed by the statistical code \textsc{Gemini} (\textsc{Gem++}
and \textsc{Gem90}) as an afterburner. 

In the literature one can find other papers \cite{Ono02,OnoJPC2013,Tian2018}
in which the results of the AMD model are compared with existing
experimental data, but they usually focus on central collisions and on
the properties of IMFs produced in light systems. 
Those papers have shown that AMD is able to reproduce in a very good way
many characteristics of such reactions.
The comparison presented in this paper demonstrates,
for the first time to our knowledge,
the capability of the AMD-plus-\textsc{Gemini} calculations
to reproduce also characteristic features of
peripheral and semiperipheral collisions in the Fermi energy regime,
where (quasi-)binary collisions still exhaust a major part of the
reaction cross section. 

Experimental data and calculated results were sorted in bins of centrality
by means of an observable, $C_b$, which is built from the secondary
velocities of QP and QT.
A good reproduction of the average velocity ratio $R_v$ and of the QP
secondary charge $Z_{QP}^{sec}$ was obtained in the calculations,
with little sensitivity to the screening parameter $y$
for the nucleon-nucleon in-medium cross section.
The QP c.m. polar angle $\theta_{\mathrm{QP}}^{\,\mathrm{cm}}$
shows some dependence on $y$, which however does not allow one to
draw definite conclusions about which value has to be preferred.

Concerning the switching time $t_{\mathrm{sw}}$ from the
dynamic code AMD to the \textsc{Gemini} afterburner,
the present analysis indicates that for the observables related to
the QP properties ($R_v$, $\theta_{QP}^{\,\mathrm{cm}}$, $Z_{QP}$) there is an
improvement when $t_{\mathrm{sw}}$ is extended at least to 500 fm/c.

The comparison between experimental data and calculations was performed
also for the total multiplicities of light charged particles and IMFs
($Z$ = 3--7) of Ref. \cite{Piantelli2006}.
It was found that the chemistry of these particles is strongly dependent
on the afterburner version that is used (\textsc{Gem90} or \textsc{Gem++}),
probably because of the different level density parametrizations in the two
versions of the code.
In fact, for a given AMD calculation, the experimental multiplicities are
better reproduced with the \textsc{Gem90} afterburner.
  Concerning the in-medium nucleon-nucleon cross section, there is no
  compelling evidence that the standard screening parameter $y$ = 0.85
  needs to be decreased in semiperipheral collisions.
The obtained agreement, although not perfect,
is reasonably acceptable for all particles, except protons, which are
always overestimated by the calculations.
However, things are not completely satisfactory and there are still 
unclear points concerning the evaporation of the \textsc{Gemini} afterburner
and its matching with the dynamic code AMD.

The calculated particle multiplicities seem generally insensitive to the
stiffness of the adopted isospin-dependent part of the equation of state.
The remarkable sensitivity of the deuteron multiplicity is most likely an
artifact due to the wrong deuteron binding energy produced by the
effective force parametrized for an asy-soft equation of state.

In conclusion, the AMD-plus-\textsc{Gemini} calculations proved to be a
reliable tool for describing heavy ion collisions in the Fermi
energy regime, not only for central collisions (as already shown
in the literature), but also for peripheral and semiperipheral ones.

\begin{acknowledgments}

  One of the authors (A. Ono) acknowledges the support from
  INFN -- Sezione di Firenze
  for his stay in 2017, and the support from Japan Society for the
  Promotion of Science KAKENHI Grant No. 17K05432 and No. 24105008.
  This work required the use of a lot of computation time for the
  production of the simulated data.
  We would like to thank the GARR Consortium for the kind use of
  the cloud computing infrastructure on the platform cloud.garr.it.
  We would like to thank also the INFN-CNAF for the use of its
  cloud computing infrastructure.

\end{acknowledgments}

\bibliography{biblio}

\end{document}